\documentclass[a4paper,11pt]{article}
\usepackage[noblocks]{authblk}
\usepackage{here}
\usepackage{subfig}
\usepackage[pdftex]{graphicx}
\usepackage{color}
\usepackage{braket}
\usepackage{amssymb, amsmath}
\usepackage{feynmf}
\usepackage{bm}
\usepackage{url}
\usepackage{slashed}
\usepackage{caption}
\usepackage{jcappub}
\usepackage[compat=1.1.0]{tikz-feynhand}
\usepackage{hyperref}
\usepackage[samesize]{cancel}
\usepackage{fancyhdr} 

\newcommand{\fng}{f_{\rm NG}}
\newcommand{\bsm}{\beta_{\rm SM}}

\newcommand{\beq}{\begin{equation}}
\newcommand{\eeq}{\end{equation}}

\begin{document}


\title{\boldmath Multi-messenger constraints on Abelian-Higgs cosmic string networks}
\author[a,b]{Mark Hindmarsh}
\author[c,d]{and Jun'ya Kume}
\affiliation[a]{Department of Physics and Helsinki Institute of Physics, PL 64, FI-00014 University of Helsinki,
Finland}
\affiliation[b]{Department of Physics and Astronomy, University of Sussex, Falmer, Brighton BN1 9QH, U.K}
\affiliation[c]{Research Center for the Early Universe (RESCEU), Graduate School of Science, The University of Tokyo, Hongo 7-3-1
Bunkyo-ku, Tokyo 113-0033, Japan}
\affiliation[d]{Department of Physics, Graduate School of Science,
The University of Tokyo, Hongo 7-3-1
Bunkyo-ku, Tokyo 113-0033, Japan}

\emailAdd{mark.hindmarsh@helsinki.fi}
\emailAdd{kjun0107@resceu.s.u-tokyo.ac.jp}

\subheader{{\rm RESCEU-15/22, HIP-2022-25/TH}}

\abstract{Nielsen-Olesen vortices in the Abelian-Higgs (AH) model are the simplest realisations of cosmic strings in a gauge field theory. 
Large-scale numerical solutions show that the dominant decay channel of a network of AH strings produced from random initial conditions is classical field radiation.
However, they also show that with special initial conditions, loops of string can be created for which classical field radiation is suppressed, and which behave like Nambu-Goto (NG) strings with a dominant decay channel into gravitational radiation.  
This indicates that cosmic strings are generically sources of both high-energy particles and gravitational waves. 
Here we adopt a simple parametrisation of the AH string network allowing for both particle and gravitational wave production. With a reference to a specific model for NG-like loop distribution, this sets the basis for a ``multi-messenger" investigation of this model. 
We find that, in order to explain the NANOGrav detection of a possible gravitational wave background, while satisfying the constraint on NG-like loop production from simulations and bounds from the cosmic microwave background, the tension of the AH string in Planck units $G\mu$ and the fraction of the NG-like loops $\fng$ should satisfy $G\mu\fng^{2.6} \gtrsim 3.2\times 10^{-13}$ at 95\% confidence.
On the other hand, for such string tensions, constraints from the diffuse gamma-ray background (DGRB) indicate that more than 97$\%$ of the total network energy should be converted to dark matter (DM) or dark radiation. 
We also consider joint constraints on the annihilation cross-section, the mass, and the relic abundance of DM produced by decays of strings. 
For example, for a DM mass of 500 GeV, the observed relic abundance can be explained by decaying AH strings that also account for the NANOGrav signal. 
}

\maketitle
\flushbottom

\section{Introduction}
Cosmic strings are linear concentrations of energy with a cosmological size that may have been produced at early-universe phase transitions in a variety of high-energy physics scenarios~\cite{Vilenkin:278400,Hindmarsh:1994re,Hindmarsh:2011qj,Copeland:2011dx}. In the traditional picture of cosmic strings, they are treated as infinitely thin line-like objects. Then the strings are expected to evolve according to the Nambu--Goto (NG) equation~\cite{Forster:1974ga,Arodz:1995dg,Anderson:1997ip} and additional re-connection rule when strings cross each other~\cite{Shellard:1987bv,Matzner:1988qqj,Achucarro:2006es}. Within this picture, an observable stochastic gravitational wave background (SGWB) can be generated by the slowly-decaying loops of cosmic string~\cite{Auclair:2019wcv}.
The most stringent constraints were provided by the non-observation of such a SGWB with the Pulsar Timing Array (PTA) experiments~\cite{Lentati:2015qwp,Shannon:2015ect,NANOGRAV:2018hou}, which gives $G\mu \lesssim 10^{-10}$ (with $\mu$ the string tension and $G$ Newton’s constant) for the pure NG-loops~\cite{Blanco-Pillado:2017rnf,Ringeval:2017eww}.
Recently, however, PTA experiments have reported the evidence of the common stochastic process in their latest results~\cite{NANOGrav:2020bcs,Goncharov:2021oub,Chen:2021rqp,Antoniadis:2022pcn}.   
Although the evidence for spatial correlations, which is indispensable to declare the SGWB detection, is as yet negligible, it is intriguing to consider the possibility that these results are associated with cosmic string~\cite{Ellis:2020ena,Blasi:2020mfx,Buchmuller:2020lbh,Blanco-Pillado:2021ygr,Bian:2020urb,Chen:2022azo,Bian:2022tju}.

In order for there to be stable string solutions, the Standard Model needs to be extended with extra spontaneously broken symmetry. 
The simplest way to extend the gauge symmetry of the Standard Model is 
with an extra U(1) symmetry, so that the new gauge field and associated scalar are described by the Abelian-Higgs (AH) model. This possesses string solutions ~\cite{Nielsen:1973cs}, which have been well investigated both analytically and numerically. 
The evolution of the AH string network is tested by the large scale numerical simulation based on the underlying classical AH field theory~\cite{Vincent:1997cx,Moore:2001px,Bevis:2006mj,Bevis:2010gj,Daverio:2015nva,Correia:2019bdl,Correia:2020gkj,Correia:2020yqg}, in which scaling behavior of the network was observed, meaning that the mean inter-string distance grows in proportion to time. 
Strings in extensions with non-Abelian symmetries behave in a similar way \cite{Hindmarsh:2018zch}.

While the observation of the scaling agrees with the modelling based on the NG approximation, AH simulations show that loops of string decay as fast as causally allowed through the production of classical scalar and gauge radiation \cite{Hindmarsh:2008dw,Hindmarsh:2021mnl}. On the basis of this fast decay, Ref.~\cite{SantanaMota:2014xpw} put bounds on the string tension from the light element abundances and the Diffuse Gamma-Ray background (DGRB), which might be affected by a chain of decay products in the visible sector from the AH strings.
If the decay products include dark matter (DM), 
further constraints follow from the dark matter relic density \cite{Hindmarsh:2013jha}.
In a recent simulation of the AH string loops~\cite{Matsunami:2019fss,Hindmarsh:2021mnl}, however, it was shown that field configurations whose evolution is well approximated by NG dynamics 
can be created by special initial conditions.
Such NG-like strings, if large enough, would decay primarily through gravitational waves. Although NG-like strings have not been observed in AH network simulations, the simulations can only put an upper bound on the fraction of such loops. Therefore we must allow for the emission of both high energy particles from the network and gravitational waves. 

This dual emission possibility motivates the investigation of possible multi-messenger signals from the AH string in this study. Following the previous work~\cite{SantanaMota:2014xpw}, we first characterize the energy injection to the visible sector from the AH string network.
As pointed out in Ref.~\cite{Vachaspati:2009kq}, the string field can generically couple to the Standard Model (SM) Higgs via the portal coupling. Therefore, a significant fraction of the network energy might end up in the SM particles, which could affect the consequence of the Big-Bang nucleosynthesis (BBN) and the DGRB measurement. 
This would also be the case for models in which the gauge symmetry of the AH model is an anomaly-free U(1) of the Standard Model such as $B-L$, where $B$ is baryon number and $L$ lepton number, such as might be expected in Grand Unified Theories \cite{Kibble:1982ae,Jeannerot:1995yn}.
Assuming that AH strings dominantly decay into SM Higgs, the authors of Ref.~\cite{SantanaMota:2014xpw} translated BBN constraints on the decaying massive particles and the upper bound on the cascade energy density into the bounds on the string tension of the AH model, which depends on this energy fraction. 
Here we update these bounds by applying the latest results on the BBN constraint ~\cite{Kawasaki:2017bqm} and the cascade energy density~\cite{Berezinsky:2016feh}.
We also briefly discuss constraint on DM production based on Ref.~\cite{Hindmarsh:2013jha}, in which the energy injection is characterized in the same way.

The intensity of the SGWB from NG-like loops in the AH model is uncertain. The authors of Ref.~\cite{Matsunami:2019fss} identified the critical length scale in their simulation above which loops are well approximated by the NG equation, and gravitational radiation presumably becomes dominant. The initial condition of the loop configurations used there, however, were somewhat special.   
In Ref.~\cite{Hindmarsh:2021mnl}, the AH loops were created from random initial conditions appropriate for cosmological network simulations~\cite{Hindmarsh:2017qff}. Long-lived NG-like loops 
were not observed in a sample of over 30 large loops, 
and the fraction of such loops was bounded at around 10\%. Following this study, we parameterize this fraction as $\fng$ to estimate the possible SGWB in the AH model. To compare it with the NANOGrav 12.5yr result~\cite{NANOGrav:2020bcs}, which assumed the power-law spectrum of the SGWB, we introduce a likelihood model and construct a mapping from the power-law parameters to the AH string parameters. Such a treatment allows us to avoid possible bias, which arises when we simply fit the SGWB spectrum by a power-law.

The rest of the paper is organized as follows. In Sec.~\ref{multi-sig}, we characterize the particle emission from the AH string network and the SGWB from NG-like loops by introducing two parameters in the model. Based on this characterization, we derive the constraints on the AH model from the optical observation and the SGWB observation in Sec.~\ref{multi-const}. We first apply the BBN constraints and the cascade energy upper bound to the decay products of the AH strings. Then we investigate the allowed parameter region of the AH model inferred from the NANOGrav 12.5yr result. By combining these two result, we derive constraints on the model parameters including the possibility of DM production from strings.
Sec.~\ref{discussion} is devoted to the discussion.

\section{Multi-messenger signals from the Abelian-Higgs string network}\label{multi-sig}
In the following, we study the AH model, whose Lagrangian density is given by
\begin{align}
    \mathcal{L} = g^{\mu\nu}D_{\mu}\phi D_{\nu}\phi^*+V(\phi)+\frac{1}{4e^2}g^{\mu\rho}g^{\nu\sigma}F_{\mu\nu}F_{\rho\sigma},
\end{align}
where $g_{\mu\nu}$ is a metric, $\phi(x)$ is a complex scalar field with the
potential $V(\phi) = \frac{1}{4}\lambda(|\phi|^2-|\phi_0|^2)$, $A_{\mu}(x)$ is a vector
field and $D_{\mu} = \partial_{\mu}-iA_{\mu}$ is the covariant derivative. 
Throughout this work, the critical coupling $\beta \equiv \lambda/2e^2 = 1$ is assumed for simplicity, where the string tension is given by $\mu = 2\pi\phi_0^2$.
We first make a brief review of the characterization of the particle radiation from the cosmic string network based on Ref.~\cite{SantanaMota:2014xpw}. Then we characterize the gravitational wave background from Nambu-Goto-like loops by utilizing the parameter advocated in Ref.~\cite{Hindmarsh:2021mnl}.
\subsection{Particle emission from the strings}
Numerical simulations of the Abelian-Higgs (AH) string networks based on the classical field theory show that loops of string quickly decay into classical radiation of the scalar and gauge field of the theory.
Since the string scalar field generically couples to the SM Higgs doublet $\Phi$ through a portal coupling, for example $\lambda_{\rm p}|\phi|^2|\Phi|^2$~\cite{Vachaspati:2009kq}, the radiation is expected to end up in the SM particles. 
This also happens if the gauge field couples to the SM through, for example, mixing with hypercharge or $B-L$.

Following the notation in Ref.~\cite{SantanaMota:2014xpw}, we parametrise the fraction of the energy which is released from the string network and results in the form of SM particles 
as $\bsm^2$. Considering covariant energy conservation, the total power per unit volume injected into the SM plasma becomes
\begin{align}
    Q_h = -\bsm^2\left[\dot{\rho}_s+3H(1+w_s)\rho_s\right],
\end{align}
where $\rho_s$ is the total energy density of the strings, $w_s$ is their average equation of state parameter and $H$ is the Hubble parameter. 
Assuming that $\rho_s$ accounts for a subdominant fraction of the total energy density of the universe $\rho$, $Q_h$ can be approximated as
\begin{align}
    \frac{Q_h}{H\rho} \simeq 3\bsm^2(w - w_s)\Omega_s,
\end{align}
where $w$ is the total equation of state parameter and $\Omega_s \equiv \rho_s/\rho$. As a result, the injected energy density in the visible sector is estimated as
\begin{align}
    \Delta\rho_h(t) \simeq Q_h t \simeq 3\bsm^2(w - w_s)\rho_s/\alpha^2, 
\end{align}
with $\alpha = \sqrt{2}$ for the radiation dominated era and $\alpha = \sqrt{3/2}$ for the matter dominated era.
In order to connect with the result of numerical simulation, let us introduce following parameters
\begin{align}
    x &\equiv \alpha\sqrt{\mu/\rho_st^2},\\
    \bar{\gamma}_{\rm SM} &\equiv 3\bsm^2(w - w_s)/x^2,
\end{align}
which allow us to rewrite $\Delta\rho_h$ in a simplified form
\begin{align}
    \Delta\rho_h(t) = \bar{\gamma}_{\rm SM}\frac{\mu}{t^2}.\label{e_inj} 
\end{align}
From the field theoretical simulation of AH model performed in Refs.~\cite{Bevis:2006mj,Hindmarsh:2008dw}, these parameters are evaluated as $x \simeq 0.7$ and $w_s = -0.13$ for the radiation era, $x \simeq 0.9$ and $w_s = -0.15$ for the matter era respectively. Combining these values, we obtain
\begin{align}
    \bar{\gamma}_{\rm SM} \simeq
\begin{cases}
2.8\bsm^2\ \ \ {\rm (RD\ era)} \\
0.5\bsm^2\ \ \ {\rm (MD\ era)}.
\end{cases}\label{gamma_ft}
\end{align}
In the following, we assume that this fraction of the classical radiation $\bsm^2$ dominantly decays into the SM Higgs (and thus we denote the energy injection as $\Delta\rho_h$). Note that the information of the coupling constant and the dynamical evolution of the particles is implicitly encoded in $\bsm$. We do not assume a specific coupling in this work and so perform a model-independent analysis by virtue of this characterization. 

This parametrisation is also applicable to the calculation of DM relic abundance as investigated in Ref.~\cite{Hindmarsh:2013jha}. We will come back to this point in the last part of Sec.~\ref{sec:comb}.

\subsection{Gravitational wave background from Nambu-Goto-like loops}\label{GWB}
If the strings obey Nambu-Goto (NG) dynamics, the SGWB from the cosmic string network is dominantly generated by the oscillation of the sub-horizon loops. Therefore, the number density ${\sf n}(l,t)$ of non-self-intersecting, sub-horizon cosmic string loops of invariant length $l$ at cosmic time $t$ is the necessary ingredient in the evaluation of the SGWB spectrum.
Interestingly, this argument might be relevant to AH string networks since the creation of the NG-like loops was shown to be possible~\cite{Matsunami:2019fss}.

While the large-scale field theory simulations in the AH model indicate that a loop of length $L$ decays into massive radiation in a time less than $0.25L$, simulations of individual loops in the AH model shows that loops whose evolution is well approximated by NG dynamics can be generated with carefully chosen initial conditions~\cite{Matsunami:2019fss,Hindmarsh:2021mnl}. This result motivates us to consider the possibility that in the AH network, a certain fraction of large loops obeys NG dynamics and survive to radiate only gravitationally~\cite{Hindmarsh:2021mnl}. 
However, such NG-like loops are yet to be observed in the numerical simulations of  networks~\cite{Hindmarsh:2021mnl}. Hence, its possible distribution is completely unknown. 
To discuss possible GW radiation from NG-like loops in AH string networks under this circumstance, we choose one of the well-established models of the non-self-intersecting loop distribution from NG simulations.

As discussed in Ref.~\cite{Auclair:2019wcv}, there are two different NG-simulation based models, the BOS model~\cite{Blanco-Pillado:2013qja, Blanco-Pillado:2017oxo} and the LRS model~\cite{Lorenz:2010sm, Ringeval:2005kr}.
The most important consequence of LRS model is the dominance of small loops, which was not observed in the BOS simulation. 
Here, in the absence of any information about the size distribution of NG-like loops in the AH simulations, 
we adopt the BOS model  
as a benchmark for the NG-like loop distribution, due to its greater simplicity. How the results of our analysis change when referring to the LRS model is briefly commented on later.
In the BOS model~\cite{Blanco-Pillado:2013qja}, the number density for non-self-intersecting loops 
is given as
\begin{align}
{\sf n}_{\rm r,r}(l,t)&=\frac{0.18}{t^{3/2}(l + \Gamma G\mu t)^{5/2}}\Theta(0.1-l/t),\label{rr}\\
{\sf n}_{\rm r,m}(l,t)&=\frac{0.18(2H_0\sqrt{\Omega_r})^{3/2}(1+z)^3}{(l + \Gamma G\mu t)^{5/2}}\Theta(0.09t_{\rm eq}/t-\Gamma G\mu -l/t),\label{rm}\\
{\sf n}_{\rm m,m}(l,t)&=\frac{0.27-0.45(l/t)^{0.31}}{t^{2}(l + \Gamma G\mu t)^{2}}\Theta(0.18-l/t),\label{mm}
\end{align}
where $H_0$ is the Hubble constant, $\Omega_{\rm r}$ is the density parameter of the radiation, $t_{\rm eq}$ represents the cosmic time of radiation-matter equality and $\Gamma$ is a constant which characterises the average total power emitted by a loop as $\Gamma G\mu^2$. Here the subscript ``r,m", for example, represents the loops produced in the radiation era and emitting GWs in matter era. 
Note that Eq.~\eqref{rm} matches to Eq.~\eqref{rr} in the early radiation era.
With this number density ${\sf n}(l,t)$, the present day spectrum of the SGWB can be calculated from
\begin{align}
\Omega_{\rm gw}^{\rm (NG)}(f) \equiv \frac{1}{\rho_c}\frac{d\ln\rho^{\rm (NG)}_{\rm gw}}{d\ln f} = \frac{8\pi f G^2\mu^2}{3H_0^2}\sum_{n = 1}^{\infty}C_n(f)P_n,    
\end{align}
where 
\begin{align}
    C_n(f)=\frac{2n}{f^2}\int_0^{\infty}\frac{dz}{H(z)(1+z)^6}{\sf n}\left(\frac{2n}{(1+z)f},t(z)\right)
\end{align}
with $H(z)$ and $t(z)$ being the Hubble parameter and cosmic time at redshift $z$.
The function $C_n(f)$ indicates the number density of loops emitting GWs observed at frequency $f$ in the $n$-th harmonic, while  
$P_n$ is the average GW power emitted by the $n$-th harmonic of a loop. 
The sum satisfies 
\begin{align}
     \sum_{n = 1}^{\infty} P_n = \Gamma.
\end{align}
The constants $P_n$ depend on the average shape of the loops and hereafter we adopt the ``smoothed" model constructed from the numerical simulation taking into account the gravitational backreaction~\cite{Blanco-Pillado:2017oxo} and take $\Gamma = 50$, which is also indicated by the simulations.


As mentioned above, distribution of the possible NG-like loops has significant uncertainty. In line with Ref.~\cite{Hindmarsh:2021mnl},
let us parameterise (a part of) this uncertainty by allowing a fraction $f_{\rm NG}$ of loops to survive to radiate only gravitationally.
Since NG-like long-lived loops were not observed in an ensemble of randomly generated AH loops, the upper bound is estimated as $\fng \lesssim 0.1$ \cite{Hindmarsh:2021mnl}.
With this parameterization, we could model the distribution of such NG-like loops in the AH network as $\fng {\sf n}(l,t)$ and thus the SGWB as
\begin{align}
\Omega_{\rm gw}^{\rm (AH)}(f) = f_{\rm NG} \Omega_{\rm gw}^{\rm (NG)}(f).\label{AHGW}  
\end{align}
We should note that use of the loop number density Eq.~\eqref{rr}--\eqref{mm} here is only a first approximation, since there is no information from the AH simulations what ${\sf n}(l,t)$ should be used. In this sense, the interpretation of $\fng$ as parametrising the fraction of NG-like loops depends on the model (BOS model in the present case).

The number density Eq.~\eqref{rr}--\eqref{mm} is based on the NG string simulation where the scaling of the infinite strings is maintained solely by the loop production at the intersection. On the other hand, scaling in AH model is also assisted by 
classical radiation. 
We therefore might expect the NG-like loops to be produced with a different initial size.  The shapes of the NG-like loops in an AH network may also be different, as the formation process is accompanied by a burst of classical radiation as the loop collapses and finds its stable configuration.  This could lead to a different gravitational wave power parameter $\Gamma$. We also neglect a next-order effect, due to particle emission at cusps and kinks, which reduces the lifetimes of smaller loops~\cite{Olum:1998ag,Matsunami:2019fss}.
This modifies the GW spectrum as investigated in Ref.~\cite{Auclair:2019jip}. For the BOS model, however, the effect of particle emission 
is to impose
a high frequency cutoff,\footnote{See also Ref.~\cite{Auclair:2021jud} which studies the effect of the particle radiation for the LRS model~\cite{Lorenz:2010sm}.} 
well above the nanohertz range of frequencies we consider.
We will leave the development of improved models to future work, with the expectation that the correction in Eq.~\eqref{AHGW} is $\mathcal{O}(1)$\footnote{For example, assuming that the loop size at the production well agrees with the BOS model, the authors of Ref.~\cite{Gouttenoire:2019kij} quantify the effect of classical radiation on the efficiency of loop production in the string network based on Ref.~\cite{Correia:2019bdl}, which results in the suppression of the overall amplitude of SGWB only by a factor $\sim2$.}.

In evaluating the gravitational wave power spectrum Eq.~\eqref{AHGW}, we assume a $\Lambda$CDM cosmology for which 
\begin{align}
    H(z) = H_0\sqrt{\Omega_{\Lambda}+(1+z)^3\Omega_m + G(z)(1+z)^4\Omega_r},
\end{align}
and take Planck 2018 fiducial parameters~\cite{Planck:2018vyg} $\Omega_\text{m} = 0.308$, $\Omega_\text{r} = 9.1476\times 10^{-5}$, $\Omega_{\Lambda} = 1 - \Omega_\text{m} - \Omega_\text{r}$ and $H_0 = 67.8$ km/s/Mpc, as adopted in Ref.~\cite{Auclair:2019wcv}. 
The change of the relativistic degrees of freedom is encoded in
\begin{align}
    \mathcal{G}(z) = \frac{g_*(z)g_s^{4/3}(0)}{g_*(0)g_s^{4/3}(z)},
\end{align}
where $g_*(z)$ and $g_s(z)$ are the effective number of relativistic and entropic degrees of freedom. To estimate it with SM particle contents, we use $g_*(T)$ and $g_s(T)$ tabulated in the micrOMEGAs~\cite{Belanger:2018ccd}, based on the calculation in Ref.~\cite{Gondolo:1990dk}.
Note that the models of loop number density~\eqref{rr}--\eqref{mm} do not take into account the variation in $\mathcal{G}(z)$. 
Therefore, we adopt an approximation described in Ref.~\cite{Blanco-Pillado:2017oxo} as follows\footnote{The effect of the variation in the degrees of freedom is analytically discussed in Ref.~\cite{Yamada:2022imq}.}.
Since in the BOS model the loop production is highly peaked at $l/t \sim 0.1$, the typical time when the loop with length $l$ at time $t$ was produced is given as
\begin{align}
    t_l = 10(l+\Gamma \mu t).
\end{align}
With this expression, the $z$ dependence of $\mathcal{G}$ can be estimated as
\begin{align}
{\sf n}_{\rm r,r}(l,t) \simeq \frac{0.18\left\{2H_0\sqrt{\Omega_r \mathcal{G}(z(t_l))}\right\}^{3/2} (1+z)^3}{(l + \Gamma G\mu t)^{5/2}}\Theta(0.1-l/t).\label{rrG}
\end{align}
Precisely speaking, this approximation is not valid when $\mathcal{G}(z(t_1))$ varies. 
However, for most values of $G\mu$ of our interest, the SGWB within PTA frequency range is dominated by the loops surviving from the radiation era~\eqref{rm} and this subtlety does not affect our result.

\begin{figure}[htbp] 
\centering
  \includegraphics[clip,width=14cm]{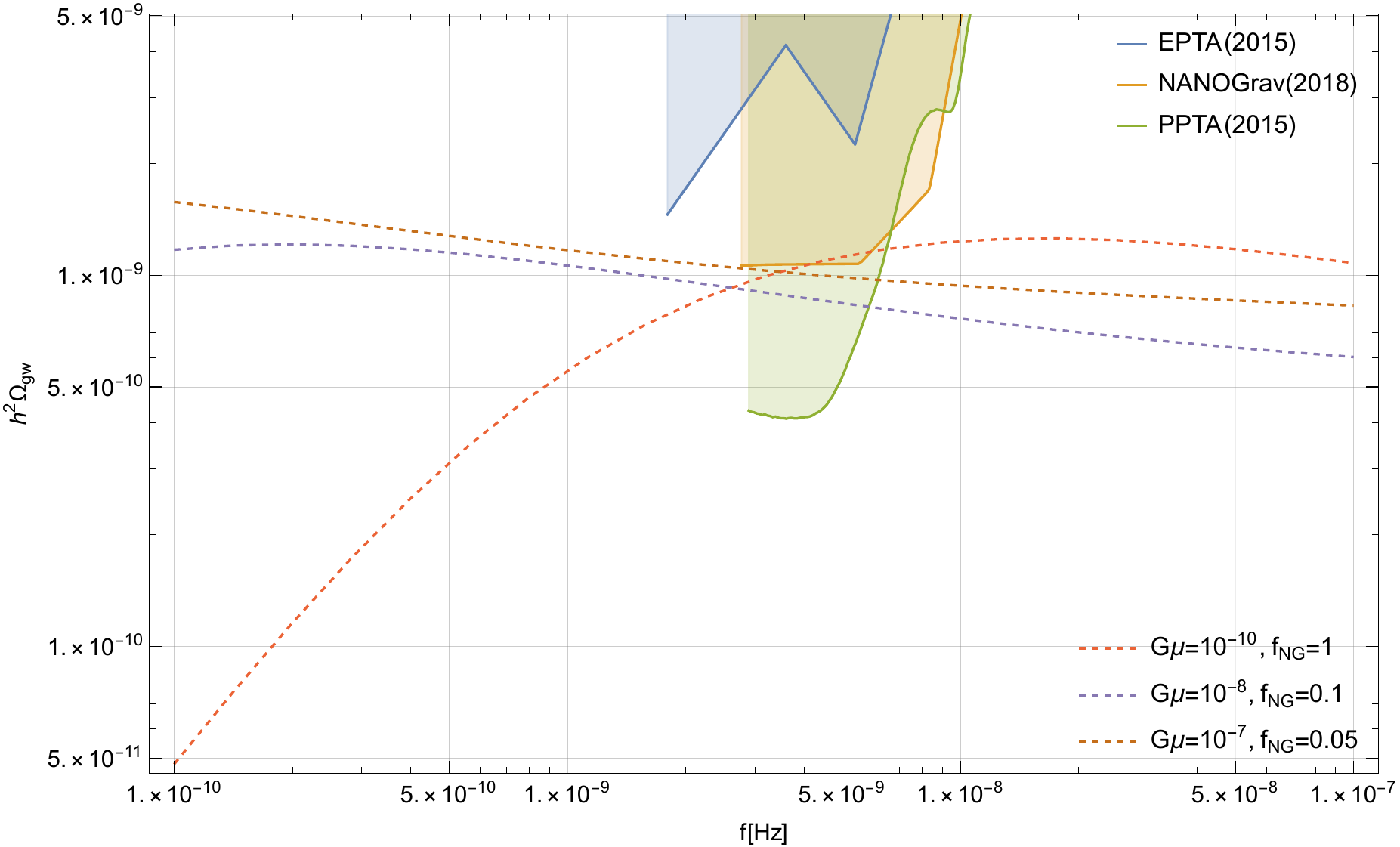}
  \caption{Examples of GWB spectra in AH model. As a reference, previous constraints by NANOGrav~\cite{NANOGRAV:2018hou}, EPTA~\cite{Lentati:2015qwp} and PPTA~\cite{Shannon:2015ect} are simultaneously plotted.
  The parameters used here are explained in the text.}
    \label{fig:gwsignal}
\end{figure}

In Fig.~\ref{fig:gwsignal}, we plot the spectrum of SGWB~\eqref{AHGW} with the benchmark parameters $(G\mu,\fng) = (10^{-10},1)$, $(10^{-8},0.1)$, and  $(10^{-7},0.05)$. 
Here we assume that the contribution from long strings are negligible, which is well justified for the parameters of our interest since it is of the order of $(G\mu)^2$ (see {\it e.g.} Ref.~\cite{CamargoNevesdaCunha:2022mvg}).
We also plot the previously obtained upper bounds by PTA experiments~\cite{NANOGRAV:2018hou,Lentati:2015qwp,Shannon:2015ect}, which is comparable to the amplitude of SGWB estimated from the recently reported common stochastic process~\cite{NANOGrav:2020bcs,Goncharov:2021oub,Chen:2021rqp,Antoniadis:2022pcn}. 
The fraction of NG-like loops $\fng = 0.1$, which is assumed for the purple line, is the highest allowed value in our model. 
On the other hand, $G\mu = 10^{-7}$ is the highest string tension which satisfies the CMB bound~\cite{Planck:2015fie,Lizarraga:2016onn}. The brown line with $\fng = 1$ is for comparison with the BOS model. We find it agrees with the spectrum depicted in the Fig.~6 of Ref.~\cite{Blanco-Pillado:2017oxo}.

The spectrum of SGWB from BOS loops has a following structure. At high frequencies, the loops radiating in the RD era yields a plateau.
At intermediate frequencies, a peak is created from GW emission from radiation-era loops in the matter era. 
Frequencies below $1/(G\mu t_0)$ are suppressed since only small number of large loops can contribute.
As investigated in Ref.~\cite{Blanco-Pillado:2021ygr}, BOS model with $G\mu \lesssim 10^{-10}$ is favoured by the recent PTA experiments. In the AH model, however, $\fng$ suppresses the overall amplitude and hence $(G\mu,\fng) = (10^{-8},0.1), (10^{-7},0.05)$ can be comparable to the BOS with $G\mu \simeq 10^{-10}$. 

\section{Multi-messenger measurements and its constraint on the AH string}\label{multi-const}

Based on the above expressions of the energy injection into the visible sector and the SGWB, here we derive constraints on the AH model by applying the NANOGrav 12.5yr result~\cite{NANOGrav:2020bcs} and the upper bound on the cascade energy density~\cite{Berezinsky:2016feh}. 
We also consider the BBN constraint on energy injection by decaying massive particles~\cite{Kawasaki:2017bqm}, finding that it is not substantially 
different from the DRGB constraint.
Then we summarize the implications for this model, including the dark matter production based on Ref.~\cite{Hindmarsh:2013jha}, obtained by combining these multi-messenger constraints.


\subsection{Big-Bang nucleosynthesis and gamma-ray constraints}\label{opt-const}
As investigated in Ref.~\cite{SantanaMota:2014xpw}, an AH string network might alter the outcomes of Big-Bang Nucleosynthesis (BBN) and contribute the diffuse gamma-ray background (DGRB) through particle emission into the cosmic medium. Here we follow the discussion in Ref.~\cite{SantanaMota:2014xpw} and update the constraints on the AH string model,
using the most recent BBN constraint on decaying massive ($\gtrsim$ GeV) particles~\cite{Kawasaki:2017bqm} and the upper bound on the cascade energy density~\cite{Berezinsky:2016feh}.

The presence of long-lived massive particles decaying into SM species 
reduces the abundance of the light elements by dissociation due to hadronic and electromagnetic showers.
Hence, the observed abundances of light elements puts bounds on the energy injection from those massive particles. In Ref.~\cite{Kawasaki:2017bqm}, the constraints are given in terms of $M_XY_X$, where $M_X$ is the mass of the new specie $X$ (or the typical energy of emitted visible particles) and $Y_X \equiv n_X/s$ is its yield at times much less than than its lifetime $\tau_X$.
As discussed in Ref.~\cite{SantanaMota:2014xpw}, this bound can be applied to the energy density injected from the AH string network in one cosmic time $\Delta\rho_h = tQ_h(t)$ by making the approximate identification $M_XY_X/\tau_X = Q_h/s$, and assuming that, for example via the emission of energetic SM Higgs, the 
SM particles produced by string decays are 
predominantly $b\bar{b}$. 
In other words, the AH string network is effectively a population of X particles decaying into $b\bar{b}$ at cosmic time $t$.
Such an identification $\tau_X \sim t$ is possible because the lifetime of unstable SM particles is much shorter than the BBN time scale.

By using Eqs.~\eqref{e_inj}--\eqref{gamma_ft} and the expression for the entropy density of the universe during RD era, 
we find
\begin{align}
     \frac{\Delta\rho_h(t)}{s(t)} \simeq 22\bsm^2G\mu\left(\frac{m_p}{t}\right)^{1/2},\label{higgs}
\end{align}
where $m_p$ is the Planck mass.
Since the dominant decay channel of the SM Higgs is $h \to b\bar{b}$, the upper right panel of the Fig.~12 in Ref.~\cite{Kawasaki:2017bqm} can be compared with Eq.~\eqref{higgs}.
Then the strongest constraint on the string tension $\mu$ comes from Deuterium at $\tau_X \sim t \simeq 3\times10^3$s as $M_XY_X \sim \Delta\rho_h(t)/s(t) \lesssim 4\times10^{-14}$GeV, which results in
\begin{align}
G\mu \lesssim 3.6\times10^{-12}\bsm^{-2}.
\end{align}
Compared to the previous work~\cite{SantanaMota:2014xpw}, the constraint becomes tighter about an order of magnitude. This tightening is due to the refinement of the BBN constraint on decaying species by adopting newer constraints on the light elements abundance and improving calculation techniques (See Ref.~\cite{Kawasaki:2017bqm} and references therein).

On the other hand, the DGRB measurement constrains the energy injection in the form of $\gamma$-rays in the late-time universe. In Ref.~\cite{Berezinsky:2016feh}, the authors derives a bound on $\omega_{\rm em}$, the total energy density of cascade at the present epoch, from the latest Fermi-LAT measurement of DGRB at GeV scale~\cite{Fermi-LAT:2014ryh}. 
This bound is given as
\begin{align}
    \omega_{\rm em} \lesssim 8.3 \times 10^{-8} {\rm eV/cm^3},
\end{align}
which is again about an order of magnitude reduction from the value used in Ref.~\cite{SantanaMota:2014xpw}.
Denoting the fraction of the energy of the string decay products that end up in $\gamma$-rays around GeV scale as $f_{\rm em}$, $\omega_{\rm em}(t)$ in the late-time universe can be expressed as
\begin{align}
    \omega_{\rm em}(t) = f_{\rm em}\Delta\rho_h(t) \simeq 0.5f_{\rm em}\bsm^2\frac{\mu}{t^2},
\end{align}
where the MD era expression of Eq.~\eqref{gamma_ft} is used for $\Delta\rho_h(t)$.
Again assuming that the decay products from strings are dominated by the SM Higgs, many photons are produced via pion decays originated from the primary decay channel of Higgs: $h\to b\bar{b}$. 
This naturally leads to $f_{\rm em} \sim 1$ and one can derive the constraint for the string tension as
\begin{align}
    G\mu \lesssim 4 \times 10^{-12}\bsm^{-2}.\label{const_DGRB}
\end{align}
Here we use only 1 significant digit because of order of magnitude estimate on the additional parameter $f_{\rm em}$. Nevertheless, $f_{\rm em} \ll 1$ should be unlikely since photons will also be produced by charged particle interactions that take place in the cosmological plasma~\cite{Bhattacharjee:1999mup}.

We should note that the both constraints depend on the details of the coupling of the string sector fields and the visible sector, but are in principle calculable once the field theory is specified.
One can see that for $\bsm^2 \lesssim 10^{-5}$, these constraints become tighter than the Planck CMB upper bound $G\mu \lesssim 10^{-7}$. 
In the following, we will use the constraint from the DGRB~\eqref{const_DGRB} to give a more conservative estimate.

\subsection{Mapping the NANOGrav 12.5yr result}\label{gwb-const}
Pulsar Timing Array experiments have recently reported strong evidence for a common-spectrum stochastic process~\cite{NANOGrav:2020bcs,Goncharov:2021oub,Chen:2021rqp,Antoniadis:2022pcn}, which might be a signal of a nanohertz SGWB. 
Here we investigate the possibility that the origin of this common stochastic process is the SGWB from the NG-like loops in the AH string network.

\begin{figure}[htbp] 
\centering
  \includegraphics[clip,width=10cm]{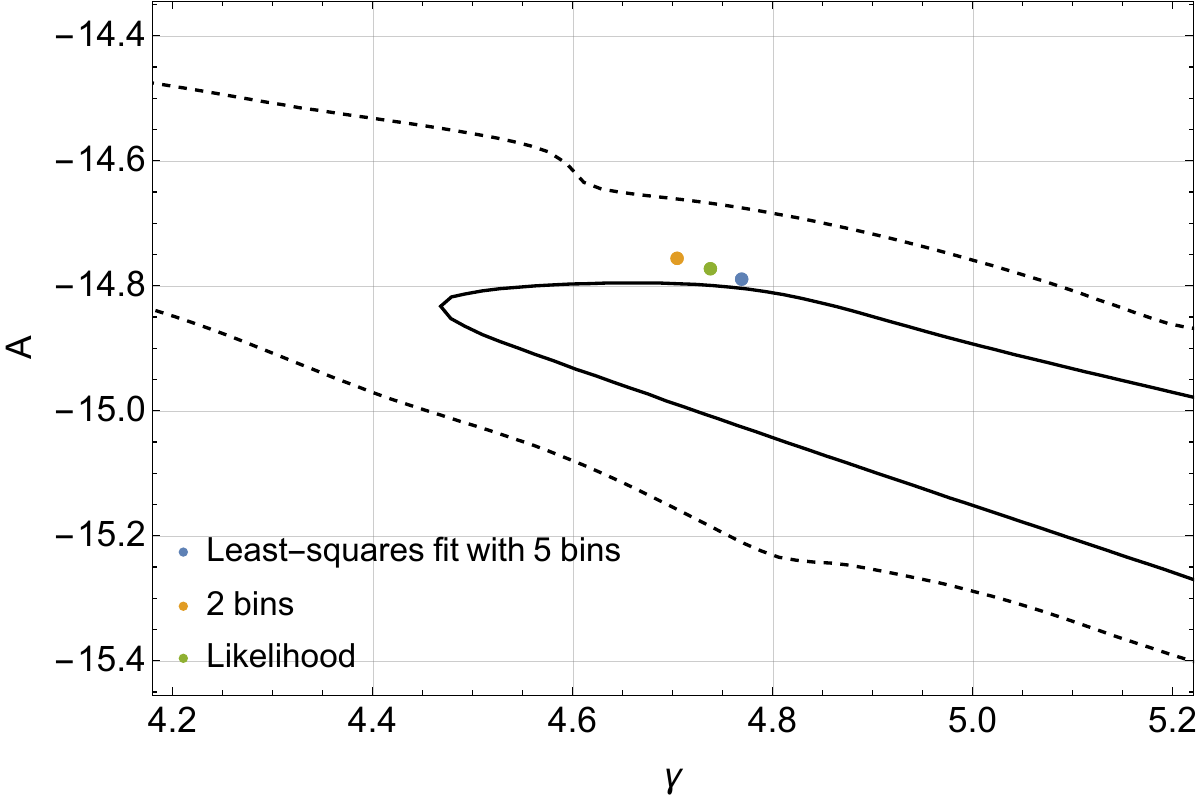}
  \caption{$68\%$ and $95\%$ posterior likelihood contours contours from NANOGrav data, and comparison between the different fitting methods with $G\mu = 10^{-10}$ and $\fng = 1$. Due to the weight taking into account the noise power, our fitting (green) lies in between the five bin numerical fitting (blue) and the lowest two bin fitting (orange).}
    \label{fig:contour_fit}
\end{figure}

In the analysis performed by the NANOGrav collaboration~\cite{NANOGrav:2020bcs}, the SGWB is modeled as a power-law around a reference frequency $f_{\rm yr} = 1$ yr as
\begin{align}
    h_c^{\rm (pow)}(f) = A\left(\frac{f}{f_{\rm yr}}\right)^{(3-\gamma)/2},
\end{align}
which is expressed in terms of the spectral GW energy density as
\begin{align}
    \Omega_{\rm gw}^{\rm (pow)}(f) = \frac{2\pi^2}{3H_0^2}f^2h_c^{\rm(pow)2}(f) = \frac{2\pi^2}{3H_0^2}A^2f_{\rm yr}^2\left(\frac{f}{f_{\rm yr}}\right)^{5-\gamma}.
\end{align}
In Fig.~\ref{fig:contour_fit}, the $68\%$ and $95\%$ posterior likelihood contours from the NANOGrav result are plotted in part of the $\gamma - A$ plane.

To connect this result with the SGWB from cosmic strings, we need construct a map between the power-law fit parameters $(\gamma, A)$ and the cosmic string parameters $(G\mu, \fng)$. In previous works~\cite{Ellis:2020ena,Blasi:2020mfx,Buchmuller:2020lbh,Blanco-Pillado:2021ygr}, this was accomplished by fitting the theoretical cosmic string spectra with power laws.
Here we adopt a method based on Ref.~\cite{Gowling:2022pzb}. 
This takes into account the fact that spectra which are not exactly power laws 
are being observed by an instrument with a frequency-dependent noise, and so 
a simple least-squares fit will be biased.
Alternatively (or more rigorously), one can perform a Bayesian analysis using NANOGrav pipeline to directly obtain a likelihood function for string-induced SGWB as in Refs.~\cite{Bian:2020urb,Chen:2022azo,Bian:2022tju}.
However, we find that our mapping seems consistent with the result obtained in Refs.~\cite{Bian:2020urb, Bian:2022tju}\footnote{The analysis in Ref.~\cite{Chen:2022azo} explicitly assumes the presence of unknown common stochastic process, which yields upper bound on the tension of the cosmic strings (or the strength of string-induced SGWB) Therefore, we cannot compare their results with ours.} and suffices our practical purpose.

Assuming Gaussian noise and the Gaussian SGWB from the AH strings, the likelihood of the data $x$ is given as
\begin{align}
    P(x|S_{\rm gw}^{\rm (AH)}) =\prod_{f_i\in\left[f_{\rm min},f_{\rm max}\right]} \frac{2T_{\rm obs}}{(2\pi)^{1/2}\left[S_n(f_i)+S_{\rm gw}^{\rm (AH)}(f_i)\right]}\exp\left(-\frac{T_{\rm obs}|\tilde{x}(f_i)|^2}{S_n(f_i)+S_{\rm gw}^{\rm (AH)}(f_i)}\right), 
\end{align}
where $S_n(f)$ is the effective strain noise spectral density and $S_{\rm gw}^{\rm (AH)}(f) = (3H_0^2/2\pi^2f^3)\Omega_{\rm gw}^{\rm (AH)}(f)$ is the spectral density of the SGWB from the cosmic strings.
In this study, we identify $S_n(f)$ as the effective sensitivity of the 12.5 yr NANOGrav estimated by a python code \texttt{Hasasia}~\cite{Hazboun2019Hasasia}, which is presented in, for example, the page 10 of Ref.~\cite{Hasasia}.
Assuming a flat prior distribution $P(S_{\rm gw}^{\rm (pow)}(A, \gamma))$, the likelihood for the power-law fitting with given data is modeled in the same way:
\begin{align}
    P(S_{\rm gw}^{\rm (pow)}|x) &\propto P(x|S_{\rm gw}^{\rm (pow)})P(S_{\rm gw}^{\rm (pow)})\notag\\
    &\propto \prod_{f_i\in\left[f_{\rm min},f_{\rm max}\right]}\frac{2T_{\rm obs}}{(2\pi)^{1/2}\left[S_n(f_i)+S_{\rm gw}^{\rm (pow)}(f_i)\right]}\exp\left(-\frac{T_{\rm obs}|\tilde{x}(f_i)|^2}{S_n(f_i)+S_{\rm gw}^{\rm (pow)}(f_i)}\right),
\end{align}
where $S_{\rm gw}^{\rm (pow)}(f) = (3H_0^2/2\pi^2f^3)\Omega_{\rm gw}^{\rm (pow)}(f)$.
By marginalizing over 
the data in frequency space, 
the likelihood function, which quantifies how probable the power-law fitting $(\gamma, A)$ is for the SGWB from the strings with a given $(G\mu,\fng)$, is constructed as
\begin{align}
    P(S_{\rm gw}^{\rm (pow)}|S_{\rm gw}^{\rm (AH)}) &= \int dx P(S_{\rm gw}^{\rm (pow)}|x)P(x|S_{\rm gw}^{\rm (AH)})\notag\\
    &\propto \prod_{f_i\in\left[f_{\rm min},f_{\rm max}\right]}\frac{\sqrt{\left[S_n(f_i)+S_{\rm gw}^{\rm (pow)}(f_i)\right]\left[S_n(f_i)+S_{\rm gw}^{\rm (AH)}(f_i)\right]}}{\left[S_n(f_i)+S_{\rm gw}^{\rm (pow)}(f_i)\right]+\left[S_n(f_i)+S_{\rm gw}^{\rm (AH)}(f_i)\right]}.\label{likelihood}
\end{align}
Each string GWB spectrum can be fitted by the power-law parameters $(\gamma, A)$ which maximize this likelihood function. 
This defines a map from $(G\mu,\fng)$ to $(\gamma, A)$.
We take five frequency bins $(f_{\rm min}, 2f_{\rm min},...,5f_{\rm min})$ in our analysis with $f_{\rm min} = 12.5$ yr, which are the same bins analyzed by the NANOGrav for the power-law fitting.

In Fig.~\ref{fig:contour_fit}, we compare our method to the simple numerical fitting and the two lowest frequency bin fitting adopted in Ref.~\cite{Blanco-Pillado:2021ygr}, for $G\mu = 10^{-10}$ and $\fng = 1$. 
The authors in Ref.~\cite{Blanco-Pillado:2021ygr} select those two bins since the noise contribution is expected to be the smallest there. This can be understood as the extreme limit of our fitting method, which is weighted by the relative size of the signal power to the modeled noise power. Thus, it is reasonable that the result of our fitting is in between the lowest two bin fitting and the numerical fitting which evenly weights the five bins. While the difference of the fitting methods seems not so significant, we use our likelihood model throughout this study to avoid the bias as much as possible.

\begin{figure}[htbp] 
\centering
  \includegraphics[clip,width=75mm]{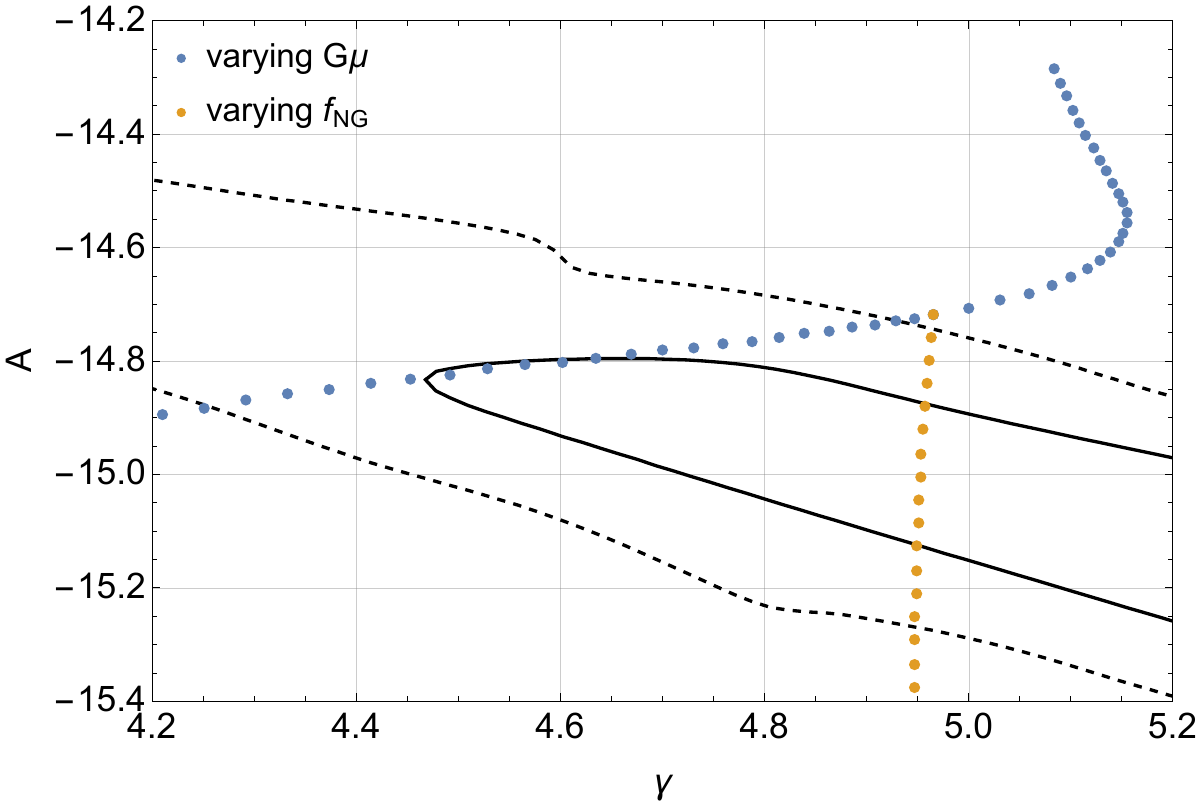}~
  \includegraphics[clip,width=75mm]{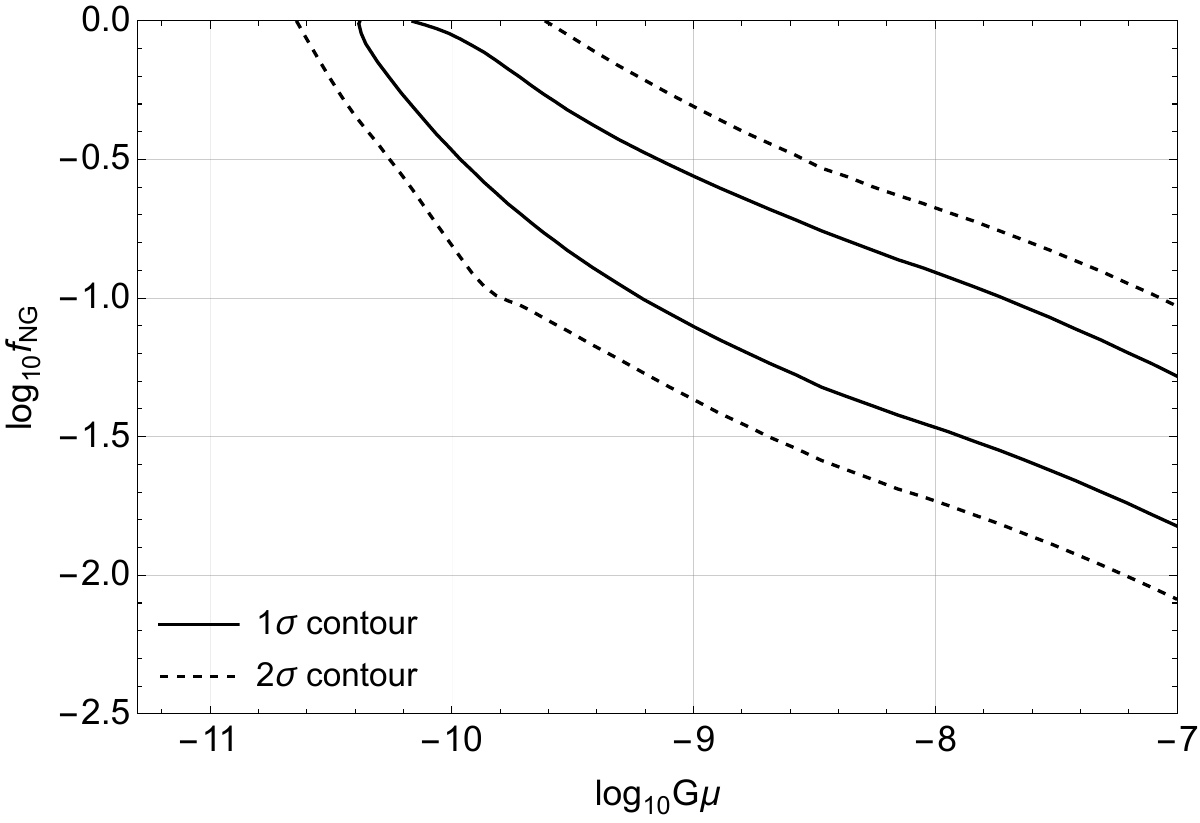}
  \caption{{\it (Left panel)} Variation of the AH string parameters $(G\mu, \fng)$ in the $\gamma - A$ plane is shown. Blue dots correspond to $\fng = 1$ and $G\mu$ is logarithmically varied from $G\mu = 10^{-7}$ to $G\mu = 3\times10^{-10}$, and from $G\mu = 3\times10^{-10}$ to $G\mu = 2\times10^{-11}$. Orange dots correspond to $G\mu = 3\times10^{-10}$ and here $\fng$ is logarythmically varied from $\fng = 1$ to $\fng = 0.07$. {\it (Right panel)} The mapping of  of the power-law fitting from the NANOGrav 12.5yr results. Here the black(-dashed) line represents contours of $68\%$ (95\%) confidence.}
    \label{fig:map}
\end{figure}

With this likelihood model, we can construct the inverse map from the space $(\gamma, A)$ to $(G\mu,\fng)$, and hence the image of NANOGrav power-law contours in the $G\mu-\fng$ plane as follows.
We first scan over the sets of string loop parameters $(G\mu,\fng)$ and find the corresponding power-law parameters $(\gamma, A)$ by maximizing the likelihood~\eqref{likelihood} as demonstrated in the left panel of Fig.~\ref{fig:map}. 
We divide the range of $G\mu$ into two sections $[2\times 10^{-11},3\times 10^{-10}]$ and $[3\times10^{-10}, 10^{-7}]$, both are sampled into 25 logarithmically spaced values. $\fng$ is also logarithmically sampled into 25 values over $10^{-2} \leq \fng \leq 1$ for each of them in the former segment, and $3\times10^{-3} \leq \fng \leq 0.3$ for the latter segment. Hence, we have $2 \times 25 \times 25$ points in total. 
With these sampling points mapped through the likelihood, the points in the $G\mu-\fng$ plane, which correspond to the points $(\gamma_c, A_c)$ on the NANOGrav contours, can be estimated.
This is the reason why we call the ``inverse map'' to $G\mu-\fng$ plane.
In this work, we use the following estimator~\cite{Gowling:2022pzb} weighted by the Euclidean distance in the mapped parameter space:
\begin{align}
    \bar{X} &= \left(\sum_{i=1}^N\frac{1}{d_i}\right)^{-1}\sum_{i=1}^N\frac{X_i}{d_i},\label{estimator}
\end{align}
where $d_i = \sqrt{(\gamma_i - \gamma_c)^2+(A_i - A_c)^2}$ is the distance between the point of interest $(\gamma_c, A_c)$ and the sampling points $(\gamma_i, A_i)$ mapped from the string parameters $X_i = (G\mu_i, f_{{\rm NG}i})$. $N$ is the number of the sampling points closest to $(\gamma_c, A_c)$, which is used in estimating the inverse map of $(\gamma_c, A_c)$, namely $\bar{X} = (\overline{G\mu},\bar{f}_{\rm NG})$. 
Note that, 
due to the characteristics of the spectrum discussed at the end of Sec.~\ref{GWB}, there can be two different values of $(G\mu,\fng)$ that are fitted to the same $(\gamma,A)$. 
In the left panel of Fig.~\ref{fig:map}, blue dots represent the mapping of points with $\fng = 1$ while $G\mu$ is varied from left (smaller) to right (larger).
Though $A$ monotonically increases with $G\mu$, $\gamma$ starts to decrease around the curve at $\gamma \simeq 5.15$ corresponding to $G\mu\simeq 5\times10^{-9}$.
On the other hand, lowering the value of $\fng$ with fixed $G\mu$ (orange dots in the left panel of Fig.~\ref{fig:map}) results in decreasing $A$ almost without changing $\gamma$ since it controls the overall amplitude of $\Omega_{\rm gw}^{\rm (AH)}$. 
From these observations, one can deduce that, by changing $\fng$, degeneracy occurs between the two branches $5\times10^{-10}\lesssim G\mu \lesssim 5\times10^{-9}$ and $5\times10^{-9}\lesssim G\mu \lesssim 10^{-7}$ reside in $\gamma \gtrsim 5$.
This indicates that we may overlook some set of the string parameters if we simply maximize the likelihood with respect to a specific $(\gamma_c, A_c)$.


The estimator \eqref{estimator} can be easily implemented to systematically obtain a correct image of the NANOGrav contours while dealing with the degeneracy mentioned above.
First, we scan the different values of $\fng$ with a fixed value of $G\mu$ as orange dots in the left panel of Fig.~\ref{fig:map} and find the intersections $(\gamma_c, A_c)$ between the interpolation of these sampling points and the contours.
Then we choose two closest sampling points $(\gamma_i, A_i)$ for each intersection to estimate the value of $(\overline{G\mu}, \bar{f}_{\rm NG})$ that would correspond to $(\gamma_c, A_c)$.
From Eq.~\eqref{estimator} with $N = 2$, we can estimate the value of $\bar{f}_{\rm NG}$ as
\begin{align}
    \bar{f}_{\rm NG} &= \left(\frac{1}{d_1}+\frac{1}{d_2}\right)^{-1}\left(\frac{f_{\rm NG1}}{d_1}+\frac{f_{\rm NG2}}{d_2}\right),\label{fng:est}
\end{align}
where $f_{\rm NG1}$ and $f_{\rm NG2}$ represent the value of $\fng$ of the two closest points, and the distance can be well approximated by $d_i \simeq |A_i - A_c|$. Note that $\overline{G\mu}$ is immediately identified with $G\mu$ since the sampling points with fixed $G\mu$ are concerned. 
We repeat this procedure for different values of $G\mu$ to obtain a set of inversely mapped points on the $G\mu - \fng$ plane. 
Importantly, this procedure performed on each value of $G\mu$ allows us to deal with parameters degenerate in $\gamma-A$ plane.
By smoothly connecting these points, the global shape of contours are depicted in the $G\mu - \fng$ plane as shown in the right panel of Fig.~\ref{fig:map}.

To refine the shape, additional points, for example, the bending of the 65\% confidence contour around $\gamma \simeq 4.5$ are also inversely mapped. In this case, the four closest sampling points were used to determine both $\overline{G\mu}$ and $\bar{f}_{\rm NG}$ through Eq.~\eqref{estimator} with $N = 4$. Note that these points are free from the degeneracy, the value of $\overline{G\mu}$ can be correctly estimated with sufficiently small spacing of the sampling points. 
As a check of the method, we confirmed that the contours in the power-law parameter space $(\gamma_c, A_c)$ are reproduced within 1\% errors by maximizing the likelihood~\eqref{likelihood} with respect to the estimated string parameters $(\overline{G\mu},\bar{f}_{\rm NG})$.

Note that the mapped contours should be regarded as an approximation of the true $68\%$ and $95\%$ contours of posterior likelihood in the $({G\mu},{f}_{\rm NG})$ space arising from the NANOGrav result.
As mentioned above, a more rigorous approach would construct the posterior with a Markov chain Monte Carlo method, sampling directly on the parameters $({G\mu},{f}_{\rm NG})$, with a physically motivated prior, using the likelihood function of the observations~\cite{Bian:2020urb,Chen:2022azo,Bian:2022tju}. 
Even without performing such runs, our method could be improved by using the chains constructed from sampling on the $(\gamma, A)$ space, and reweighting according to the prior induced by the mapping between the two parameter spaces~\cite{Gowling:2022pzb}.
Nevertheless, our present results, $10^{-10.4} \lesssim G\mu \lesssim 10^{-10.2}$ for $\fng = 1$ at 68\% confidence well overlap with $\log_{10}G\mu = -10.38^{+0.21}_{-0.21}$ obtained in Ref.~\cite{Bian:2022tju} where pure BOS model was analyzed (see Table I therein)\footnote{We believe that the slight discrepancy is due to our approximate method and to the fact that the SGWB spectra used in Ref.~\cite{Bian:2022tju} (see Fig.~1 therein) differ somewhat from those of Ref.~\cite{Blanco-Pillado:2017oxo} and ours.}. 
Therefore, the mapped contour in the right panel of Fig.~\ref{fig:map} can be consistently used in the following discussion.

\subsection{Combined constraints and implications for the AH model}\label{sec:comb}

\begin{figure}[htbp] 
\centering
  \includegraphics[clip,width=14cm]{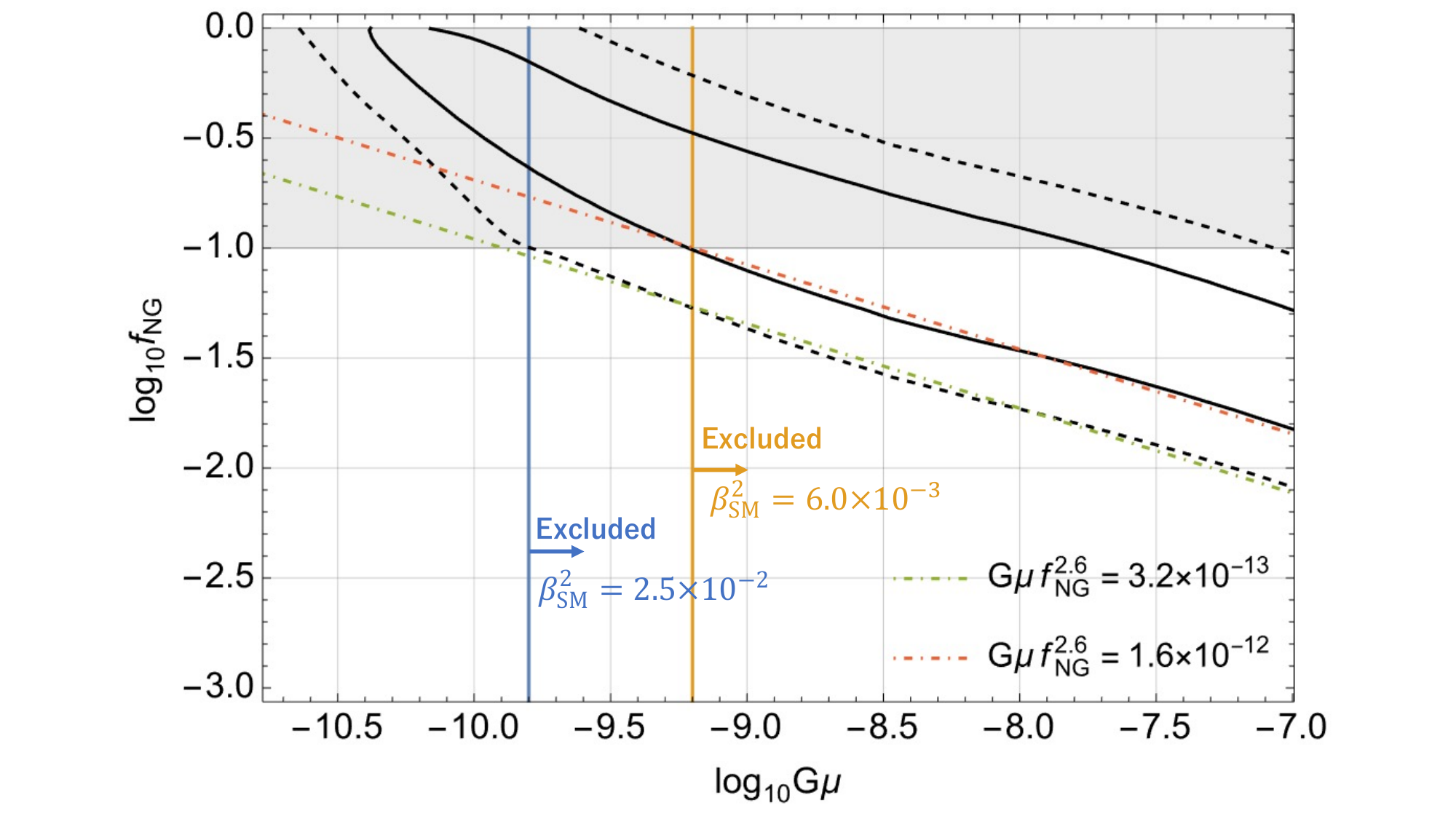}
  \caption{Constraints on AH string model parameters which allow an explanation of the NANOGrav result, while satisfying constraints from numerical simulations and the DGRB. 
  The mapped NANOGrav $68\%$ and $95\%$ likelihood contours are shown as solid black and dashed black lines. The lower branch of each contour is approximated by the dot-dashed lines with constant $G\mu\fng^{2.6}$ in the region $\fng \lesssim 0.1$ and $G\mu \lesssim 10^{-7}$. The region excluded by the upper bound on the fraction of NG-like loops $\fng \lesssim 0.1$ from numerical simulations \cite{Hindmarsh:2021mnl} is indicated by grey shading. 
  The blue and orange solid lines correspond to the lowest values of $G\mu$ required to explain the NANOGrav result that are within the $65\%$ and $95\%$ confidence regions. The annotation with the same color illustrates the DGRB constraint~\eqref{const_DGRB} on the fractional energy of the stirngs going into Standard Model particles $\bsm^2$, where those values of $G\mu$ become the maximally allowed value. 
  }\label{fig:comb_const}
\end{figure}

Here we combine the constraints from the NANOGrav 12.5 yr data and the DGRB observations by Fermi-LAT with the Planck CMB upper bound  $G\mu \lesssim 10^{-7}$ and the the constraint $\fng \lesssim 0.1$ from simulations to investigate the allowed parameter region of the AH model~\cite{Planck:2015fie,Lizarraga:2016onn}. 
These constraints are summarised in Fig.~\ref{fig:comb_const}.
In the region $\fng \lesssim 0.1$ and $G\mu \lesssim 10^{-7}$, the lower branches of the likelihood contours from NANOGrav can be approximately characterised as
\begin{equation}
    G\mu \fng^{2.6} \gtrsim 3.2 \times 10^{-13} \; (95\% ), \quad G\mu \fng^{2.6} \gtrsim 1.6 \times 10^{-12} \; (68\% ),\label{eq:contour_const}
\end{equation}
which are plotted as the colored dot-dashed line.
Note that if we instead use the LRS model as a reference, this constraint should be modified because of the differences in the SGWB spectra.
Such a difference should be evident especially for $G\mu \gtrsim 10^{-10}$ where the difference in spectra in both amplitude and slope in the nanohertz band become relevant.
Our naive expectation is that since LRS model typically predicts larger amplitude than that of BOS model due to the dominance of small loops~\cite{Auclair:2019wcv}, preferred values of $\fng$ there tends to be smaller for LRS model, which reproduce the similar order of intensity of the SGWB as BOS model.

From Fig.~\ref{fig:comb_const}, one can see that the AH model should have $\fng \gtrsim 0.01$ in order for there to exist string tensions within the 95\% confidence region which satisfy the CMB constraint. This bound is a factor of 10 larger than the rough estimate in the previous study~\cite{Hindmarsh:2021mnl}, due to our more careful analysis.
Such a lower bound on the value of $\fng$ provides an experimentally motivated target for the numerical simulation of the AH string loops.

In addition, the string field needs to be sufficiently ``dark", or sequestered from the visible sector, if AH strings are to explain the NANOGrav result. 
The colored vertical lines in Fig.~\ref{fig:comb_const} indicates the lowest values of the string tension within the 
$65\%$ and $95\%$ confidence regions of parameter space accounting for the NANOGrav result, where the exclusion of $\fng \gtrsim 0.1$ 
is taken into account.
In order for such values of string tension to satisfy the DRGB constraint~\eqref{const_DGRB}, in which the maximally allowed value of $G\mu$ monotonically increases with decreasing $\bsm$, $\bsm$ needs to be bounded from above.
A string tension $\log_{10}G\mu \gtrsim -9.8$ is required to reach the 95\% confidence region, which in turn requires $\bsm^2 \lesssim 3 \times 10^{-2}$ in order for AH strings to satisfy the DRGB bound while accounting for the NANOGrav signal.
At $65\%$ confidence, a string tension $\log_{10}G\mu \gtrsim -9.2$ is required, which implies $\bsm^2 \lesssim 6\times 10^{-3}$.

Let us further discuss the implications of our result for the AH model. The decay products of the massive radiation from the AH string network should end up distributed between SM particles, dark matter and dark radiation. If we characterize this branching ratio of the energy density as $\beta_{\rm DM}^2$ and $\beta_{\rm DR}^2$, in the same way as $\bsm^2$, $\bsm^2 + \beta_{\rm DM}^2 + \beta_{\rm DR}^2 \lesssim 1$ should be satisfied.\footnote{Note that this sum is slightly less than 1 due to the GW emission from the NG-like loops for the values of $\fng$ of our interest.}
As discussed in the above, $\bsm^2 < 0.03$, or $\beta_{\rm DM}^2 + \beta_{\rm DR}^2 > 0.97$ is required for the AH string to explain NANOGrav result. 
While the decay into the dark radiation cannot be tightly constrained, the dark matter (DM) production from the strings can give a constraint on the new string parameters. 

For example, Ref.~\cite{Hindmarsh:2013jha} considers production of TeV scale DM $\chi$ from decaying cosmic strings via the intermediate state $X$. There its mass was assumed as $m_{\chi} = 500$ GeV for the concreteness and we also assume it in the following.
The constraints are summarized in the Table 2 of Ref.~\cite{Hindmarsh:2013jha} and here we briefly discuss its application.  
Assuming the standard freeze-out scenario, it was shown that cosmic strings in the AH model (referred to as the field theory scenario) can be a principal source of DM. 
In this case, the relic abundance depends on the energy injection characterized by $G\mu\beta_{\rm DM}^2$ in the same way as\footnote{Here we can effectively identify $\beta_{\rm DM}^2$ with $f_X$ in Ref.~\cite{Hindmarsh:2013jha}. Then $P_{\rm FT}$ in Ref.~\cite{Hindmarsh:2013jha} can be expressed as $P_{\rm FT} \simeq 8.9\times10^{-2}\beta_{\rm DM}^2$ for RD era.}  Eqs.~\eqref{e_inj}--\eqref{gamma_ft} and on $\sigma_0 \equiv \langle\sigma_{\chi} v\rangle|_{T = m_{\chi}}$, which is the thermally averaged DM annihilation cross section at the temperature equal to DM mass $m_{\chi}$.
Hence, depending on whether s-wave or p-wave annihilation is dominant, relic DM abundance measured by Planck gives the upper bound on this injection and hence a bound on $G\mu$ as 
\begin{align}
    G\mu  \lesssim
\begin{cases}
1\times10^{-9}\beta_{\rm DM}^{-2}\left(\sigma_0/10^{-23}{\rm cm^3s^{-1}}\right) & {\rm (s\mathchar`-wave)} \\
7\times10^{-12}\beta_{\rm DM}^{-2}\left(\sigma_0/10^{-23}{\rm cm^3s^{-1}}\right)^{1/2} & {\rm (p\mathchar`-wave)},\label{DMGmu}
\end{cases}
\end{align}
where equality corresponds to the case where observed amount of DM is reproduced by the string decay. 
To account for the relic DM abundance simultaneously for the NANOGrav signal, which favours larger value of $G\mu$, smaller values of $\beta_{\rm DM}$ and $\sigma_0$ might be required from Eq.~\eqref{DMGmu}.
However, one cannot increase the annihilation cross section arbitrarily due to unitarity~\cite{Griest:1989wd}.
This unitarity bound on the cross section can be recast for our reference value of $m_\chi = 500$ GeV as 
\begin{align}
    G\mu \lesssim 
\begin{cases}
2\times10^{-7}\beta_{\rm DM}^{-2} & {\rm (s\mathchar`-wave)} \\
8\times10^{-10}\beta_{\rm DM}^{-2} & {\rm (p\mathchar`-wave)},
\end{cases}
\end{align}
where equality in Eq.~\eqref{DMGmu} is assumed. 
One can quickly establish that 
accounting for the NANOGrav result with a larger $G\mu$ does lead to a constraint on the DM decay fraction for p-wave dominance.
For example, $G\mu \simeq 10^{-8}$ can yield the correct DM relic abundance without violating p-wave unitarity only if $\beta_{\rm DM}^2 \lesssim 0.1$. 
In this case, to avoid the DGRB constraint, most of the AH string decay products should then end up in the form of dark radiation as $\beta_{\rm DR}^2 \gtrsim 0.9$.

For s-wave dominance, a constraint from the Fermi-LAT observation of $\gamma$-ray in dwarf spheroidal satellite galaxies~\cite{Fermi-LAT:2011vow} is also considered in Ref.~\cite{Hindmarsh:2013jha}. This gives $G\mu \lesssim 10^{-10}\beta_{\rm DM}^{-2}$ and 
again smaller $\beta^2_{\rm DM}$ (and larger $\beta^2_{\rm DR}$ to avoid the DGRB constraint) is required for AH strings which account for the NANOGrav signal with a larger $G\mu$.

In this way, the branching ratio of the AH string decay products can be investigated for a specific model.
Although it generally depends on details of the DM, such constraints on this ratio have an important implication to how the AH model should be embedded in the model of particle physics.
The intriguing possibility that strings in a particular model could account for significant fractions of the NANOGrav signal, the Fermi-LAT measurement of the DRGB, and the dark matter is still open.

\section{Discussion}\label{discussion}
In this study, we investigated the Abelian-Higgs model in light of recent results of PTA experiments which can be interpreted in terms of a stochastic gravitational wave background. Motivated by the recent simulation of the AH string loops
~\cite{Matsunami:2019fss, Hindmarsh:2021mnl}, we consider the possibility that AH strings radiate both particles and gravitational waves. By introducing parameters $\fng$, $\bsm^2$, $\beta_{\rm DM}^2$, and $\beta_{\rm DR}^2$ we characterize these two classes of signals from the AH string network. 
The parameter $\fng$ accounts for the uncertainty in the understanding of the NG-like loop production in the AH model, 
while the parameters $\bsm^2$, $\beta_{\rm DM}^2$, and $\beta_{\rm DR}^2$ are the model-dependent fractional energies of the string network appearing as Standard Model particles, dark matter, and dark radiation.  
The decay fraction $\bsm^2$ is bounded by the diffuse $\gamma$-ray background observations of Fermi-LAT and by the bounds on dissociation of light elements during Big Bang nucleosynthesis. 

We modelled the uncertainty by supposing that a fraction of the AH string loops are NG-like and follow 
the loop distribution inferred from NG simulations~\cite{Blanco-Pillado:2013qja}, 
leading to a SGWB a factor $\fng$ smaller than the NG prediction. 
We should note that our prediction highly depends on the assumed model of loop distribution. Therefore, further understanding of the NG-like loop production is indispensable for the refinement of our multi-messenger investigation.  
Resultant improvement of the SGWB estimation from the AH string network may allow us to precisely probe AH model by the multi-band GWB observation including ground-based detectors and space-based detectors.

Then we probed the allowed parameter region of the AH strings by combining the constraints on gravitational wave production from  NANOGrav~\cite{NANOGrav:2020bcs} with updated constraints on particle production from the DGRB and BBN. In practice, the particle production constraints are very similar and we used only the DGRB.
In contrast to the previous studies on the cosmic string interpretation of the NANOGrav result~\cite{Ellis:2020ena,Blasi:2020mfx,Buchmuller:2020lbh,Blanco-Pillado:2021ygr}, which simply fitted the SGWB from the cosmic strings by a power-law, we model the likelihood function for the AH string parameters, which has the advantage of avoiding a possible bias in the posterior.
By maximizing this likelihood function, we map the sets of string parameters to the power-law parameter space. By inverting the map, we reconstruct 
the NANOGrav likelihood contours in the string parameter space.
We find that, to account for the NANOGrav result with approximately 95\% confidence, while respecting the CMB bound on the string tension $G\mu \lesssim 10^{-7}$ and the simulation bound on the NG-like loop fraction $\fng \lesssim 0.1$, the combination $G\mu \fng^{2.6} \gtrsim 3.2 \times 10^{-13} \; (95\% )$ is required. From this, the DGRB bound on $G\mu$ implies an upper bound on SM decay fraction as $\bsm^2 < 0.03$,
which indicates that more than 97$\%$ of the total network energy should end up in the form of dark matter or dark radiation.
Interestingly, there is a possibility that the observed relic DM abundance can be explained by a specific AH model satisfying all the constraints mentioned above.  

Due to the suppression of the GWB amplitude by $\fng$, a higher value of the string tension is still viable for AH strings and the relatively high scale symmetry-breaking might be possible, up to the CMB bound $G\mu \lesssim 10^{-7}$. 
If the fraction of NG-like loops is sufficiently low, 
$f_{\rm NG} \lesssim 0.01$, the NANOGrav result cannot be explained by AH strings.  Establishing whether the upper limit on $\fng$ is lower than this value motivates large-scale numerical simulation of AH loop production. 

For AH strings which do account for the NANOGrav signal, the resulting bound on the decay fraction into SM particles $\bsm^2 < 0.03$ 
motivates the exploration of models where the string fields are weakly coupled to the SM. In the context of such SM extensions, 
one may further discuss decays into dark matter and dark radiation, and the constraints on $\beta_{\rm DM}$ and $\beta_{\rm DR}$.
In this way, multi-messenger investigation can provide information SM extensions with spontaneously broken U(1) symmetries.
However, since the evidence of SGWB detection is not conclusive in the present PTA results, we should carefully follow future results and analysis.

The extension of our study can be considered in the following directions. First, relation between the decay fraction parameters $\bsm, \beta_{\rm DM(R)}$ and $\fng$ can be investigated. While we have treated them independently, these parameters can be related to each other through the scaling of the network which is maintained by the particle emission and NG-like loop production. This will require a clearer understanding of loop production in AH model.
Second, it will be interesting to translate the DGRB and BBN bounds into the constraints on the parameters in the Lagrangian of specific models.
This motivates investigation of the dependence of $\bsm$ and $\beta_{\rm DM(R)}$ on coupling constants.
It will be also interesting to investigate other models of cosmic string in this multi-messenger context.
For example, our results are particularly relevant for $B-L$ strings 
\cite{Jeannerot:1995yn,Sahu:2004ir,Buchmuller:2013lra}, although 
all the constraints we consider can be avoided if the U(1) symmetry is embedded in a larger symmetry group so that the strings are metastable \cite{Buchmuller:2019gfy}. In such models strong SGWBs can be produced at early times, and be observable at LISA and other instruments \cite{Buchmuller:2019gfy}, without violating any of the constraints discussed here.

\section*{Acknowledgments}
The authors would like to thank Takeo Moroi for useful comments on the BBN constraint.
JK (ORCID ID 0000-0003-3126-5100) is supported by JSPS KAKENHI, Grant-in-Aid for JSPS Fellows 20J21866 and research program of the Leading Graduate Course for Frontiers of Mathematical Sciences and Physics (FMSP). JK thanks the Helsinki Institute of Physics for hosting his visit to conduct this project and also for their hospitality.
MH (ORCID ID 0000-0002-9307-437X) acknowledges support from the Academy of Finland grant no.~333609. 

\bibliographystyle{JHEP}
\bibliography{main}

\providecommand{\href}[2]{#2}\begingroup\raggedright\begin{thebibliography}{10}

\bibitem{Vilenkin:278400}
A.~Vilenkin and E.~P.~S. Shellard, \emph{{Cosmic strings and other topological
  defects}}.
\newblock Cambridge monographs on mathematical physics. Cambridge Univ. Press,
  Cambridge, 1994.

\bibitem{Hindmarsh:1994re}
M.~B. Hindmarsh and T.~W.~B. Kibble, \emph{{Cosmic strings}},
  \href{https://doi.org/10.1088/0034-4885/58/5/001}{\emph{Rept. Prog. Phys.}
  {\bfseries 58} (1995) 477--562},
  [\href{https://arxiv.org/abs/hep-ph/9411342}{{\ttfamily hep-ph/9411342}}].

\bibitem{Hindmarsh:2011qj}
M.~Hindmarsh, \emph{{Signals of Inflationary Models with Cosmic Strings}},
  \href{https://doi.org/10.1143/PTPS.190.197}{\emph{Prog. Theor. Phys. Suppl.}
  {\bfseries 190} (2011) 197--228},
  [\href{https://arxiv.org/abs/1106.0391}{{\ttfamily 1106.0391}}].

\bibitem{Copeland:2011dx}
E.~J. Copeland, L.~Pogosian and T.~Vachaspati, \emph{{Seeking String Theory in
  the Cosmos}},
  \href{https://doi.org/10.1088/0264-9381/28/20/204009}{\emph{Class. Quant.
  Grav.} {\bfseries 28} (2011) 204009},
  [\href{https://arxiv.org/abs/1105.0207}{{\ttfamily 1105.0207}}].

\bibitem{Forster:1974ga}
D.~Forster, \emph{{Dynamics of Relativistic Vortex Lines and their Relation to
  Dual Theory}},
  \href{https://doi.org/10.1016/0550-3213(74)90008-X}{\emph{Nucl. Phys. B}
  {\bfseries 81} (1974) 84--92}.

\bibitem{Arodz:1995dg}
H.~Arodz, \emph{{Expansion in the width: The case of vortices}},
  \href{https://doi.org/10.1016/0550-3213(95)00386-7}{\emph{Nucl. Phys. B}
  {\bfseries 450} (1995) 189--208},
  [\href{https://arxiv.org/abs/hep-th/9503001}{{\ttfamily hep-th/9503001}}].

\bibitem{Anderson:1997ip}
M.~R. Anderson, F.~Bonjour, R.~Gregory and J.~Stewart, \emph{{Effective action
  and motion of a cosmic string}},
  \href{https://doi.org/10.1103/PhysRevD.56.8014}{\emph{Phys. Rev. D}
  {\bfseries 56} (1997) 8014--8028},
  [\href{https://arxiv.org/abs/hep-ph/9707324}{{\ttfamily hep-ph/9707324}}].

\bibitem{Shellard:1987bv}
E.~P.~S. Shellard, \emph{{Cosmic String Interactions}},
  \href{https://doi.org/10.1016/0550-3213(87)90290-2}{\emph{Nucl. Phys. B}
  {\bfseries 283} (1987) 624--656}.

\bibitem{Matzner:1988qqj}
R.~A. Matzner, \emph{{Interaction of U(1) cosmic strings: Numerical
  intercommutation}}, \href{https://doi.org/10.1063/1.168306}{\emph{Comput.
  Phys.} {\bfseries 2} (1988) 51--64}.

\bibitem{Achucarro:2006es}
A.~Achucarro and R.~de~Putter, \emph{{Effective non-intercommutation of local
  cosmic strings at high collision speeds}},
  \href{https://doi.org/10.1103/PhysRevD.74.121701}{\emph{Phys. Rev. D}
  {\bfseries 74} (2006) 121701},
  [\href{https://arxiv.org/abs/hep-th/0605084}{{\ttfamily hep-th/0605084}}].

\bibitem{Auclair:2019wcv}
P.~Auclair et~al., \emph{{Probing the gravitational wave background from cosmic
  strings with LISA}},
  \href{https://doi.org/10.1088/1475-7516/2020/04/034}{\emph{JCAP} {\bfseries
  04} (2020) 034}, [\href{https://arxiv.org/abs/1909.00819}{{\ttfamily
  1909.00819}}].

\bibitem{Lentati:2015qwp}
L.~Lentati et~al., \emph{{European Pulsar Timing Array Limits On An Isotropic
  Stochastic Gravitational-Wave Background}},
  \href{https://doi.org/10.1093/mnras/stv1538}{\emph{Mon. Not. Roy. Astron.
  Soc.} {\bfseries 453} (2015) 2576--2598},
  [\href{https://arxiv.org/abs/1504.03692}{{\ttfamily 1504.03692}}].

\bibitem{Shannon:2015ect}
R.~M. Shannon et~al., \emph{{Gravitational waves from binary supermassive black
  holes missing in pulsar observations}},
  \href{https://doi.org/10.1126/science.aab1910}{\emph{Science} {\bfseries 349}
  (2015) 1522--1525}, [\href{https://arxiv.org/abs/1509.07320}{{\ttfamily
  1509.07320}}].

\bibitem{NANOGRAV:2018hou}
{\scshape NANOGRAV} collaboration, Z.~Arzoumanian et~al., \emph{{The NANOGrav
  11-year Data Set: Pulsar-timing Constraints On The Stochastic
  Gravitational-wave Background}},
  \href{https://doi.org/10.3847/1538-4357/aabd3b}{\emph{Astrophys. J.}
  {\bfseries 859} (2018) 47},
  [\href{https://arxiv.org/abs/1801.02617}{{\ttfamily 1801.02617}}].

\bibitem{Blanco-Pillado:2017rnf}
J.~J. Blanco-Pillado, K.~D. Olum and X.~Siemens, \emph{{New limits on cosmic
  strings from gravitational wave observation}},
  \href{https://doi.org/10.1016/j.physletb.2018.01.050}{\emph{Phys. Lett. B}
  {\bfseries 778} (2018) 392--396},
  [\href{https://arxiv.org/abs/1709.02434}{{\ttfamily 1709.02434}}].

\bibitem{Ringeval:2017eww}
C.~Ringeval and T.~Suyama, \emph{{Stochastic gravitational waves from cosmic
  string loops in scaling}},
  \href{https://doi.org/10.1088/1475-7516/2017/12/027}{\emph{JCAP} {\bfseries
  12} (2017) 027}, [\href{https://arxiv.org/abs/1709.03845}{{\ttfamily
  1709.03845}}].

\bibitem{NANOGrav:2020bcs}
{\scshape NANOGrav} collaboration, Z.~Arzoumanian et~al., \emph{{The NANOGrav
  12.5 yr Data Set: Search for an Isotropic Stochastic Gravitational-wave
  Background}},
  \href{https://doi.org/10.3847/2041-8213/abd401}{\emph{Astrophys. J. Lett.}
  {\bfseries 905} (2020) L34},
  [\href{https://arxiv.org/abs/2009.04496}{{\ttfamily 2009.04496}}].

\bibitem{Goncharov:2021oub}
B.~Goncharov et~al., \emph{{On the Evidence for a Common-spectrum Process in
  the Search for the Nanohertz Gravitational-wave Background with the Parkes
  Pulsar Timing Array}},
  \href{https://doi.org/10.3847/2041-8213/ac17f4}{\emph{Astrophys. J. Lett.}
  {\bfseries 917} (2021) L19},
  [\href{https://arxiv.org/abs/2107.12112}{{\ttfamily 2107.12112}}].

\bibitem{Chen:2021rqp}
S.~Chen et~al., \emph{{Common-red-signal analysis with 24-yr high-precision
  timing of the European Pulsar Timing Array: inferences in the stochastic
  gravitational-wave background search}},
  \href{https://doi.org/10.1093/mnras/stab2833}{\emph{Mon. Not. Roy. Astron.
  Soc.} {\bfseries 508} (2021) 4970--4993},
  [\href{https://arxiv.org/abs/2110.13184}{{\ttfamily 2110.13184}}].

\bibitem{Antoniadis:2022pcn}
J.~Antoniadis et~al., \emph{{The International Pulsar Timing Array second data
  release: Search for an isotropic gravitational wave background}},
  \href{https://doi.org/10.1093/mnras/stab3418}{\emph{Mon. Not. Roy. Astron.
  Soc.} {\bfseries 510} (2022) 4873--4887},
  [\href{https://arxiv.org/abs/2201.03980}{{\ttfamily 2201.03980}}].

\bibitem{Ellis:2020ena}
J.~Ellis and M.~Lewicki, \emph{{Cosmic String Interpretation of NANOGrav Pulsar
  Timing Data}},
  \href{https://doi.org/10.1103/PhysRevLett.126.041304}{\emph{Phys. Rev. Lett.}
  {\bfseries 126} (2021) 041304},
  [\href{https://arxiv.org/abs/2009.06555}{{\ttfamily 2009.06555}}].

\bibitem{Blasi:2020mfx}
S.~Blasi, V.~Brdar and K.~Schmitz, \emph{{Has NANOGrav found first evidence for
  cosmic strings?}},
  \href{https://doi.org/10.1103/PhysRevLett.126.041305}{\emph{Phys. Rev. Lett.}
  {\bfseries 126} (2021) 041305},
  [\href{https://arxiv.org/abs/2009.06607}{{\ttfamily 2009.06607}}].

\bibitem{Buchmuller:2020lbh}
W.~Buchmuller, V.~Domcke and K.~Schmitz, \emph{{From NANOGrav to LIGO with
  metastable cosmic strings}},
  \href{https://doi.org/10.1016/j.physletb.2020.135914}{\emph{Phys. Lett. B}
  {\bfseries 811} (2020) 135914},
  [\href{https://arxiv.org/abs/2009.10649}{{\ttfamily 2009.10649}}].

\bibitem{Blanco-Pillado:2021ygr}
J.~J. Blanco-Pillado, K.~D. Olum and J.~M. Wachter, \emph{{Comparison of cosmic
  string and superstring models to NANOGrav 12.5-year results}},
  \href{https://doi.org/10.1103/PhysRevD.103.103512}{\emph{Phys. Rev. D}
  {\bfseries 103} (2021) 103512},
  [\href{https://arxiv.org/abs/2102.08194}{{\ttfamily 2102.08194}}].

\bibitem{Bian:2020urb}
L.~Bian, R.-G. Cai, J.~Liu, X.-Y. Yang and R.~Zhou, \emph{{Evidence for
  different gravitational-wave sources in the NANOGrav dataset}},
  \href{https://doi.org/10.1103/PhysRevD.103.L081301}{\emph{Phys. Rev. D}
  {\bfseries 103} (2021) L081301},
  [\href{https://arxiv.org/abs/2009.13893}{{\ttfamily 2009.13893}}].

\bibitem{Chen:2022azo}
Z.-C. Chen, Y.-M. Wu and Q.-G. Huang, \emph{{Search for the Gravitational-wave
  Background from Cosmic Strings with the Parkes Pulsar Timing Array Second
  Data Release}},  \href{https://arxiv.org/abs/2205.07194}{{\ttfamily
  2205.07194}}.

\bibitem{Bian:2022tju}
L.~Bian, J.~Shu, B.~Wang, Q.~Yuan and J.~Zong, \emph{{Searching for cosmic
  string induced stochastic gravitational wave background with the Parkes
  Pulsar Timing Array}},  \href{https://arxiv.org/abs/2205.07293}{{\ttfamily
  2205.07293}}.

\bibitem{Nielsen:1973cs}
H.~B. Nielsen and P.~Olesen, \emph{{Vortex Line Models for Dual Strings}},
  \href{https://doi.org/10.1016/0550-3213(73)90350-7}{\emph{Nucl. Phys. B}
  {\bfseries 61} (1973) 45--61}.

\bibitem{Vincent:1997cx}
G.~Vincent, N.~D. Antunes and M.~Hindmarsh, \emph{{Numerical simulations of
  string networks in the Abelian Higgs model}},
  \href{https://doi.org/10.1103/PhysRevLett.80.2277}{\emph{Phys. Rev. Lett.}
  {\bfseries 80} (1998) 2277--2280},
  [\href{https://arxiv.org/abs/hep-ph/9708427}{{\ttfamily hep-ph/9708427}}].

\bibitem{Moore:2001px}
J.~N. Moore, E.~P.~S. Shellard and C.~J. A.~P. Martins, \emph{{On the evolution
  of Abelian-Higgs string networks}},
  \href{https://doi.org/10.1103/PhysRevD.65.023503}{\emph{Phys. Rev. D}
  {\bfseries 65} (2002) 023503},
  [\href{https://arxiv.org/abs/hep-ph/0107171}{{\ttfamily hep-ph/0107171}}].

\bibitem{Bevis:2006mj}
N.~Bevis, M.~Hindmarsh, M.~Kunz and J.~Urrestilla, \emph{{CMB power spectrum
  contribution from cosmic strings using field-evolution simulations of the
  Abelian Higgs model}},
  \href{https://doi.org/10.1103/PhysRevD.75.065015}{\emph{Phys. Rev. D}
  {\bfseries 75} (2007) 065015},
  [\href{https://arxiv.org/abs/astro-ph/0605018}{{\ttfamily
  astro-ph/0605018}}].

\bibitem{Bevis:2010gj}
N.~Bevis, M.~Hindmarsh, M.~Kunz and J.~Urrestilla, \emph{{CMB power spectra
  from cosmic strings: predictions for the Planck satellite and beyond}},
  \href{https://doi.org/10.1103/PhysRevD.82.065004}{\emph{Phys. Rev. D}
  {\bfseries 82} (2010) 065004},
  [\href{https://arxiv.org/abs/1005.2663}{{\ttfamily 1005.2663}}].

\bibitem{Daverio:2015nva}
D.~Daverio, M.~Hindmarsh, M.~Kunz, J.~Lizarraga and J.~Urrestilla,
  \emph{{Energy-momentum correlations for Abelian Higgs cosmic strings}},
  \href{https://doi.org/10.1103/PhysRevD.95.049903}{\emph{Phys. Rev. D}
  {\bfseries 93} (2016) 085014},
  [\href{https://arxiv.org/abs/1510.05006}{{\ttfamily 1510.05006}}].

\bibitem{Correia:2019bdl}
J.~R. C. C.~C. Correia and C.~J. A.~P. Martins, \emph{{Extending and
  Calibrating the Velocity dependent One-Scale model for Cosmic Strings with
  One Thousand Field Theory Simulations}},
  \href{https://doi.org/10.1103/PhysRevD.100.103517}{\emph{Phys. Rev. D}
  {\bfseries 100} (2019) 103517},
  [\href{https://arxiv.org/abs/1911.03163}{{\ttfamily 1911.03163}}].

\bibitem{Correia:2020gkj}
J.~R. C. C.~C. Correia and C.~J. A.~P. Martins, \emph{{Quantifying the effect
  of cooled initial conditions on cosmic string network evolution}},
  \href{https://doi.org/10.1103/PhysRevD.102.043503}{\emph{Phys. Rev. D}
  {\bfseries 102} (2020) 043503},
  [\href{https://arxiv.org/abs/2007.12008}{{\ttfamily 2007.12008}}].

\bibitem{Correia:2020yqg}
J.~R. C. C.~C. Correia and C.~J. A.~P. Martins,
  \emph{{Abelian\textendash{}Higgs cosmic string evolution with multiple
  GPUs}}, \href{https://doi.org/10.1016/j.ascom.2020.100438}{\emph{Astron.
  Comput.} {\bfseries 34} (2021) 100438},
  [\href{https://arxiv.org/abs/2005.14454}{{\ttfamily 2005.14454}}].

\bibitem{Hindmarsh:2018zch}
M.~Hindmarsh, A.~Kormu, A.~Lopez-Eiguren and D.~J. Weir, \emph{{Scaling in
  necklaces of monopoles and semipoles}},
  \href{https://doi.org/10.1103/PhysRevD.98.103533}{\emph{Phys. Rev. D}
  {\bfseries 98} (2018) 103533},
  [\href{https://arxiv.org/abs/1809.03384}{{\ttfamily 1809.03384}}].

\bibitem{Hindmarsh:2008dw}
M.~Hindmarsh, S.~Stuckey and N.~Bevis, \emph{{Abelian Higgs Cosmic Strings:
  Small Scale Structure and Loops}},
  \href{https://doi.org/10.1103/PhysRevD.79.123504}{\emph{Phys. Rev. D}
  {\bfseries 79} (2009) 123504},
  [\href{https://arxiv.org/abs/0812.1929}{{\ttfamily 0812.1929}}].

\bibitem{Hindmarsh:2021mnl}
M.~Hindmarsh, J.~Lizarraga, A.~Urio and J.~Urrestilla, \emph{{Loop decay in
  Abelian-Higgs string networks}},
  \href{https://doi.org/10.1103/PhysRevD.104.043519}{\emph{Phys. Rev. D}
  {\bfseries 104} (2021) 043519},
  [\href{https://arxiv.org/abs/2103.16248}{{\ttfamily 2103.16248}}].

\bibitem{SantanaMota:2014xpw}
H.~F. Santana~Mota and M.~Hindmarsh, \emph{{Big-Bang Nucleosynthesis and
  Gamma-Ray Constraints on Cosmic Strings with a large Higgs condensate}},
  \href{https://doi.org/10.1103/PhysRevD.91.043001}{\emph{Phys. Rev. D}
  {\bfseries 91} (2015) 043001},
  [\href{https://arxiv.org/abs/1407.3599}{{\ttfamily 1407.3599}}].

\bibitem{Hindmarsh:2013jha}
M.~Hindmarsh, R.~Kirk and S.~M. West, \emph{{Dark Matter from Decaying
  Topological Defects}},
  \href{https://doi.org/10.1088/1475-7516/2014/03/037}{\emph{JCAP} {\bfseries
  03} (2014) 037}, [\href{https://arxiv.org/abs/1311.1637}{{\ttfamily
  1311.1637}}].

\bibitem{Matsunami:2019fss}
D.~Matsunami, L.~Pogosian, A.~Saurabh and T.~Vachaspati, \emph{{Decay of Cosmic
  String Loops Due to Particle Radiation}},
  \href{https://doi.org/10.1103/PhysRevLett.122.201301}{\emph{Phys. Rev. Lett.}
  {\bfseries 122} (2019) 201301},
  [\href{https://arxiv.org/abs/1903.05102}{{\ttfamily 1903.05102}}].

\bibitem{Vachaspati:2009kq}
T.~Vachaspati, \emph{{Cosmic Rays from Cosmic Strings with Condensates}},
  \href{https://doi.org/10.1103/PhysRevD.81.043531}{\emph{Phys. Rev. D}
  {\bfseries 81} (2010) 043531},
  [\href{https://arxiv.org/abs/0911.2655}{{\ttfamily 0911.2655}}].

\bibitem{Kibble:1982ae}
T.~W.~B. Kibble, G.~Lazarides and Q.~Shafi, \emph{{Strings in SO(10)}},
  \href{https://doi.org/10.1016/0370-2693(82)90829-2}{\emph{Phys. Lett. B}
  {\bfseries 113} (1982) 237--239}.

\bibitem{Jeannerot:1995yn}
R.~Jeannerot, \emph{{A Supersymmetric SO(10) model with inflation and cosmic
  strings}}, \href{https://doi.org/10.1103/PhysRevD.53.5426}{\emph{Phys. Rev.
  D} {\bfseries 53} (1996) 5426--5436},
  [\href{https://arxiv.org/abs/hep-ph/9509365}{{\ttfamily hep-ph/9509365}}].

\bibitem{Kawasaki:2017bqm}
M.~Kawasaki, K.~Kohri, T.~Moroi and Y.~Takaesu, \emph{{Revisiting Big-Bang
  Nucleosynthesis Constraints on Long-Lived Decaying Particles}},
  \href{https://doi.org/10.1103/PhysRevD.97.023502}{\emph{Phys. Rev. D}
  {\bfseries 97} (2018) 023502},
  [\href{https://arxiv.org/abs/1709.01211}{{\ttfamily 1709.01211}}].

\bibitem{Berezinsky:2016feh}
V.~Berezinsky and O.~Kalashev, \emph{{High energy electromagnetic cascades in
  extragalactic space: physics and features}},
  \href{https://doi.org/10.1103/PhysRevD.94.023007}{\emph{Phys. Rev. D}
  {\bfseries 94} (2016) 023007},
  [\href{https://arxiv.org/abs/1603.03989}{{\ttfamily 1603.03989}}].

\bibitem{Hindmarsh:2017qff}
M.~Hindmarsh, J.~Lizarraga, J.~Urrestilla, D.~Daverio and M.~Kunz,
  \emph{{Scaling from gauge and scalar radiation in Abelian Higgs string
  networks}}, \href{https://doi.org/10.1103/PhysRevD.96.023525}{\emph{Phys.
  Rev. D} {\bfseries 96} (2017) 023525},
  [\href{https://arxiv.org/abs/1703.06696}{{\ttfamily 1703.06696}}].

\bibitem{Blanco-Pillado:2013qja}
J.~J. Blanco-Pillado, K.~D. Olum and B.~Shlaer, \emph{{The number of cosmic
  string loops}}, \href{https://doi.org/10.1103/PhysRevD.89.023512}{\emph{Phys.
  Rev. D} {\bfseries 89} (2014) 023512},
  [\href{https://arxiv.org/abs/1309.6637}{{\ttfamily 1309.6637}}].

\bibitem{Blanco-Pillado:2017oxo}
J.~J. Blanco-Pillado and K.~D. Olum, \emph{{Stochastic gravitational wave
  background from smoothed cosmic string loops}},
  \href{https://doi.org/10.1103/PhysRevD.96.104046}{\emph{Phys. Rev. D}
  {\bfseries 96} (2017) 104046},
  [\href{https://arxiv.org/abs/1709.02693}{{\ttfamily 1709.02693}}].

\bibitem{Lorenz:2010sm}
L.~Lorenz, C.~Ringeval and M.~Sakellariadou, \emph{{Cosmic string loop
  distribution on all length scales and at any redshift}},
  \href{https://doi.org/10.1088/1475-7516/2010/10/003}{\emph{JCAP} {\bfseries
  10} (2010) 003}, [\href{https://arxiv.org/abs/1006.0931}{{\ttfamily
  1006.0931}}].

\bibitem{Ringeval:2005kr}
C.~Ringeval, M.~Sakellariadou and F.~Bouchet, \emph{{Cosmological evolution of
  cosmic string loops}},
  \href{https://doi.org/10.1088/1475-7516/2007/02/023}{\emph{JCAP} {\bfseries
  02} (2007) 023}, [\href{https://arxiv.org/abs/astro-ph/0511646}{{\ttfamily
  astro-ph/0511646}}].

\bibitem{Olum:1998ag}
K.~D. Olum and J.~J. Blanco-Pillado, \emph{{Field theory simulation of Abelian
  Higgs cosmic string cusps}},
  \href{https://doi.org/10.1103/PhysRevD.60.023503}{\emph{Phys. Rev. D}
  {\bfseries 60} (1999) 023503},
  [\href{https://arxiv.org/abs/gr-qc/9812040}{{\ttfamily gr-qc/9812040}}].

\bibitem{Auclair:2019jip}
P.~Auclair, D.~A. Steer and T.~Vachaspati, \emph{{Particle emission and
  gravitational radiation from cosmic strings: observational constraints}},
  \href{https://doi.org/10.1103/PhysRevD.101.083511}{\emph{Phys. Rev. D}
  {\bfseries 101} (2020) 083511},
  [\href{https://arxiv.org/abs/1911.12066}{{\ttfamily 1911.12066}}].

\bibitem{Auclair:2021jud}
P.~Auclair, K.~Leyde and D.~A. Steer, \emph{{A window for cosmic strings}},
  \href{https://arxiv.org/abs/2112.11093}{{\ttfamily 2112.11093}}.

\bibitem{Gouttenoire:2019kij}
Y.~Gouttenoire, G.~Servant and P.~Simakachorn, \emph{{Beyond the Standard
  Models with Cosmic Strings}},
  \href{https://doi.org/10.1088/1475-7516/2020/07/032}{\emph{JCAP} {\bfseries
  07} (2020) 032}, [\href{https://arxiv.org/abs/1912.02569}{{\ttfamily
  1912.02569}}].

\bibitem{Planck:2018vyg}
{\scshape Planck} collaboration, N.~Aghanim et~al., \emph{{Planck 2018 results.
  VI. Cosmological parameters}},
  \href{https://doi.org/10.1051/0004-6361/201833910}{\emph{Astron. Astrophys.}
  {\bfseries 641} (2020) A6},
  [\href{https://arxiv.org/abs/1807.06209}{{\ttfamily 1807.06209}}].

\bibitem{Belanger:2018ccd}
G.~B\'elanger, F.~Boudjema, A.~Goudelis, A.~Pukhov and B.~Zaldivar,
  \emph{{micrOMEGAs5.0 : Freeze-in}},
  \href{https://doi.org/10.1016/j.cpc.2018.04.027}{\emph{Comput. Phys. Commun.}
  {\bfseries 231} (2018) 173--186},
  [\href{https://arxiv.org/abs/1801.03509}{{\ttfamily 1801.03509}}].

\bibitem{Gondolo:1990dk}
P.~Gondolo and G.~Gelmini, \emph{{Cosmic abundances of stable particles:
  Improved analysis}},
  \href{https://doi.org/10.1016/0550-3213(91)90438-4}{\emph{Nucl. Phys. B}
  {\bfseries 360} (1991) 145--179}.

\bibitem{Yamada:2022imq}
M.~Yamada and K.~Yonekura, \emph{{Cosmic strings from pure Yang-Mills theory}},
   \href{https://arxiv.org/abs/2204.13123}{{\ttfamily 2204.13123}}.

\bibitem{CamargoNevesdaCunha:2022mvg}
D.~Camargo Neves~da Cunha, C.~Ringeval and F.~R. Bouchet, \emph{{Stochastic
  gravitational waves from long cosmic strings}},
  \href{https://doi.org/10.1088/1475-7516/2022/09/078}{\emph{JCAP} {\bfseries
  09} (2022) 078}, [\href{https://arxiv.org/abs/2205.04349}{{\ttfamily
  2205.04349}}].

\bibitem{Planck:2015fie}
{\scshape Planck} collaboration, P.~A.~R. Ade et~al., \emph{{Planck 2015
  results. XIII. Cosmological parameters}},
  \href{https://doi.org/10.1051/0004-6361/201525830}{\emph{Astron. Astrophys.}
  {\bfseries 594} (2016) A13},
  [\href{https://arxiv.org/abs/1502.01589}{{\ttfamily 1502.01589}}].

\bibitem{Lizarraga:2016onn}
J.~Lizarraga, J.~Urrestilla, D.~Daverio, M.~Hindmarsh and M.~Kunz, \emph{{New
  CMB constraints for Abelian Higgs cosmic strings}},
  \href{https://doi.org/10.1088/1475-7516/2016/10/042}{\emph{JCAP} {\bfseries
  10} (2016) 042}, [\href{https://arxiv.org/abs/1609.03386}{{\ttfamily
  1609.03386}}].

\bibitem{Fermi-LAT:2014ryh}
{\scshape Fermi-LAT} collaboration, M.~Ackermann et~al., \emph{{The spectrum of
  isotropic diffuse gamma-ray emission between 100 MeV and 820 GeV}},
  \href{https://doi.org/10.1088/0004-637X/799/1/86}{\emph{Astrophys. J.}
  {\bfseries 799} (2015) 86},
  [\href{https://arxiv.org/abs/1410.3696}{{\ttfamily 1410.3696}}].

\bibitem{Bhattacharjee:1999mup}
P.~Bhattacharjee and G.~Sigl, \emph{{Origin and propagation of extremely
  high-energy cosmic rays}},
  \href{https://doi.org/10.1016/S0370-1573(99)00101-5}{\emph{Phys. Rept.}
  {\bfseries 327} (2000) 109--247},
  [\href{https://arxiv.org/abs/astro-ph/9811011}{{\ttfamily
  astro-ph/9811011}}].

\bibitem{Gowling:2022pzb}
C.~Gowling, M.~Hindmarsh, D.~C. Hooper and J.~Torrado, \emph{{Reconstructing
  physical parameters from template gravitational wave spectra at LISA: first
  order phase transitions}},
  \href{https://arxiv.org/abs/2209.13551}{{\ttfamily 2209.13551}}.

\bibitem{Hazboun2019Hasasia}
J.~Hazboun, J.~Romano and T.~Smith, \emph{Hasasia: A python package for pulsar
  timing array sensitivity curves},
  \href{https://doi.org/10.21105/joss.01775}{\emph{Journal of Open Source
  Software} {\bfseries 4} (10, 2019) 1775}.

\bibitem{Hasasia}
J.~Hazboun, ``{Pulsar Timing Array GW Astronomy Update}.''
  \textsc{url:}~\url{https://dcc.ligo.org/public/0177/G2101389/001/jsh_gwanw_2021.pdf},
  2021.

\bibitem{Griest:1989wd}
K.~Griest and M.~Kamionkowski, \emph{{Unitarity Limits on the Mass and Radius
  of Dark Matter Particles}},
  \href{https://doi.org/10.1103/PhysRevLett.64.615}{\emph{Phys. Rev. Lett.}
  {\bfseries 64} (1990) 615}.

\bibitem{Fermi-LAT:2011vow}
{\scshape Fermi-LAT} collaboration, M.~Ackermann et~al., \emph{{Constraining
  Dark Matter Models from a Combined Analysis of Milky Way Satellites with the
  Fermi Large Area Telescope}},
  \href{https://doi.org/10.1103/PhysRevLett.107.241302}{\emph{Phys. Rev. Lett.}
  {\bfseries 107} (2011) 241302},
  [\href{https://arxiv.org/abs/1108.3546}{{\ttfamily 1108.3546}}].

\bibitem{Sahu:2004ir}
N.~Sahu, P.~Bhattacharjee and U.~A. Yajnik, \emph{{B - L cosmic strings and
  baryogenesis}}, \href{https://doi.org/10.1103/PhysRevD.70.083534}{\emph{Phys.
  Rev. D} {\bfseries 70} (2004) 083534},
  [\href{https://arxiv.org/abs/hep-ph/0406054}{{\ttfamily hep-ph/0406054}}].

\bibitem{Buchmuller:2013lra}
W.~Buchm\"uller, V.~Domcke, K.~Kamada and K.~Schmitz, \emph{{The Gravitational
  Wave Spectrum from Cosmological $B-L$ Breaking}},
  \href{https://doi.org/10.1088/1475-7516/2013/10/003}{\emph{JCAP} {\bfseries
  10} (2013) 003}, [\href{https://arxiv.org/abs/1305.3392}{{\ttfamily
  1305.3392}}].

\bibitem{Buchmuller:2019gfy}
W.~Buchmuller, V.~Domcke, H.~Murayama and K.~Schmitz, \emph{{Probing the scale
  of grand unification with gravitational waves}},
  \href{https://doi.org/10.1016/j.physletb.2020.135764}{\emph{Phys. Lett. B}
  {\bfseries 809} (2020) 135764},
  [\href{https://arxiv.org/abs/1912.03695}{{\ttfamily 1912.03695}}].

\end{thebibliography}\endgroup
\end{document}